\documentclass[fleqn,usenatbib]{mnras}

\usepackage{newtxtext,newtxmath}

\usepackage[T1]{fontenc}

\DeclareRobustCommand{\VAN}[3]{#2}
\let\VANthebibliography\thebibliography
\def\thebibliography{\DeclareRobustCommand{\VAN}[3]{##3}\VANthebibliography}

\usepackage{graphicx}	
\usepackage{amsmath}	
\usepackage[separate-uncertainty=true,multi-part-units=single]{siunitx}
\usepackage{multicol}
\usepackage{multirow}
\usepackage{xcolor}
\usepackage{placeins}

\DeclareMathOperator\erf{erf}

\newcommand{\cthree}{\textsuperscript{13}{\rm C}}
\newcommand{\ctwo}{\textsuperscript{12}{\rm C}}
\newcommand{\cratio}{\textsuperscript{12}C/\textsuperscript{13}C}
\newcommand{\invcratio}{\textsuperscript{13}C/\textsuperscript{12}C}
\newcommand{\cratiomath}{{\ensuremath{\rm \textsuperscript{12}C / \textsuperscript{13}C}}}
\newcommand{\invcratiomath}{{\ensuremath{\rm \textsuperscript{13}C / \textsuperscript{12}C}}}

\newcommand{\zem}{\ensuremath{z_{\rm em}}}
\newcommand{\zabs}{\ensuremath{z_{\rm abs}}}

\newcommand{\atom}[2]{#1{\sc \,#2}}

\newcommand{\abundance}[2]{\ensuremath{\left[{\rm #1/#2}\right]}}

\DeclareSIUnit\ADU{ADU}
\DeclareSIUnit\Angstrom{\textup{\AA}}
\newcommand{\kmps}[1]{\qty[retain-explicit-plus]{#1}{\kilo\meter\per\second}}

\newcommand{\upl}{\texttt{UVES\_popler}}
\newcommand{\aivpfit}{\texttt{AI-VPFIT}}

\newcommand{\tabref}[1]{Table \ref{#1}}
\newcommand{\figref}[1]{Figure \ref{#1}}
\newcommand{\secref}[1]{Section \ref{#1}}

\title[\textsuperscript{12}C/\textsuperscript{13}C in the DLA 1331$+$170]{ Isotopic abundance of carbon in the DLA towards QSO B1331$+$170}

\author[Milakovi\'c et al.]{
Dinko Milakovi\'c\textsuperscript{1,2}\thanks{E-mail: dinko@milakovic.net},
John K.\ Webb\textsuperscript{3,4,5},
Paolo Molaro\textsuperscript{2,1},
Chung-Chi Lee\textsuperscript{5},
Prashin Jethwa\textsuperscript{6},\newauthor
Guido Cupani\textsuperscript{2,1},
Michael T.\ Murphy\textsuperscript{7}, 
Louise Welsh\textsuperscript{2,1},
Valentina D'Odorico\textsuperscript{2,1},
Stefano Cristiani\textsuperscript{2,1,8},\newauthor
Ricardo G{\'e}nova Santos\textsuperscript{9,10},
Carlos J.\ A.\ P.\ Martins\textsuperscript{11,12}, 
Nelson J.\ Nunes\textsuperscript{13,14},
Tobias M.\ Schmidt\textsuperscript{15},\newauthor
Francesco A.\ Pepe\textsuperscript{15},
Maria Rosa Zapatero Osorio\textsuperscript{16},
Yann Alibert\textsuperscript{17,18}, 
J.\ I.\ Gonz\'alez Hern\'andez\textsuperscript{9,10},\newauthor
Paolo Di Marcantonio\textsuperscript{2},
Enric Palle\textsuperscript{9,10},
Rafael Rebolo\textsuperscript{9},
Nuno C.\ Santos\textsuperscript{12,19},
S\'ergio G.\ Sousa\textsuperscript{12}, \newauthor and
Alejandro Su{\'a}rez Mascare{\~n}o\textsuperscript{9,10}.
\\
\textsuperscript{1} Institute for Fundamental Physics of the Universe, Via Beirut, 2, 34151 Trieste, Italy\\
\textsuperscript{2} INAF - Osservatorio Astronomico di Trieste, via Tiepolo 11, 34131, Trieste, Italy\\
\textsuperscript{3} Clare Hall, University of Cambridge, Herschel Rd, Cambridge CB3 9AL, UK\\
\textsuperscript{4} Institute of Astronomy, Madingley Road, Cambridge, CB3 0HA, UK\\
\textsuperscript{5} Big Questions Institute, Level 4, 55 Holt St., Surry Hills, Sydney, NSW 2010, Australia\\
\textsuperscript{6} University of Vienna, Department of Astrophysics, T\"urkenschanzstraße 17, 1180, Vienna, Austria\\
\textsuperscript{7} Centre for Astrophysics and Supercomputing, Swinburne University of Technology, Hawthorn, Victoria 3122, Australia\\
\textsuperscript{8} INFN - National Institute for Nuclear Physics, via Valerio 2, I-34127 Trieste\\
\textsuperscript{9} Instituto de Astrof\'{\i}sica de Canarias, V\'ia L\'actea s/n, E-38200 La Laguna, Tenerife, Spain\\
\textsuperscript{10} Departamento de Astrof\'{\i}sica, Universidad de La Laguna, Avenida Astrof\'{\i}sico Francisco S\'anchez s/n, E-38206 La Laguna, Tenerife, Spain\\
\textsuperscript{11} {Centro de Astrof\'isica da Universidade do Porto, Rua das Estrelas, 4150-762 Porto, Portugal}\\
\textsuperscript{12} {Instituto de Astrof\'{\i}sica e Ci\^encias do Espa\c{c}o, CAUP, Rua das Estrelas, 4150-762 Porto, Portugal}\\
\textsuperscript{13} Instituto de Astrof\'isica e Ciências do Espa\c{c}o, Faculdade de Ci\^encias da Universidade de Lisboa, Campo Grande, PT1749-016 Lisboa, Portugal\\
\textsuperscript{14} Departamento de Fisica da Faculdade de Cincias da Universidade de Lisboa, Edifcio C8, 1749-016 Lisboa, Portugal\\
\textsuperscript{15} Observatoire Astronomique de l'Université de Genève, Chemin Pegasi 51, CH-1290 Versoix, Switzerland\\
\textsuperscript{16} Centro de Astrobiología (CSIC-INTA), Crta. Ajalvir km 4, E-28850 Torrejón de Ardoz, Madrid, Spain\\
\textsuperscript{17} Physikalisches Institut, Space Division, Universität Bern, Gesselschaftsstrasse 6, 3012 Bern, Switzerland\\
\textsuperscript{18} Center for Space and Habitability, Universität Bern, Gesselschaftsstrasse 6, 3012 Bern, Switzerland\\
\textsuperscript{19} Departamento de F\'isica e Astronomia, Faculdade de Ci\^encias, Universidade do Porto, Rua do Campo Alegre, 4169-007 Porto, Portugal
}

\date{Accepted XXX. Received YYY; in original form ZZZ}

\pubyear{2015}

\begin{document}
\label{firstpage}
\pagerange{\pageref{firstpage}--\pageref{lastpage}}
\maketitle

\begin{abstract}
Chemical evolution models predict a gradual build-up of {\cthree} in the universe, based on empirical nuclear reaction rates and assumptions on the properties of stellar populations. However, old metal-poor stars within the Galaxy contain more {\cthree} than is predicted, suggesting that further refinements to the models are necessary. Gas at high redshift provides important supplementary information at metallicities $-2\lesssim \abundance{Fe}{H}\lesssim-1$, for which there are only a few measurements in the Galaxy. We obtained new, high-quality, VLT/ESPRESSO observations of the QSO B1331$+$170 and used them to measure {\cratio} in the damped Lyman-$\alpha$ system (DLA) at $\zabs=1.776$, with \abundance{Fe}{H}=-1.27. {\aivpfit}, an Artificial Intelligence tool based on genetic algorithms and guided by a spectroscopic information criterion, was used to explore different possible kinematic structures of the carbon gas. Three hundred independent {\aivpfit} models of the absorption system were produced using pre-set {\cratio} values, ranging from 4 to 500. Our results show that $\cratiomath=28.5^{+51.5}_{-10.4}$, suggesting a possibility of {\cthree} production at low metallicity. 
\end{abstract}




\section{Introduction}
Galaxy formation and chemical evolution models are guided by the observed chemical element abundances in stars of different metallicities and in diffuse gas.
Isotopic ratio measurements provide additional constraints, since different isotopes are produced by different nuclear reactions within stellar interiors. Given the similar atomic structure of isotopes, measuring their relative abundances is relatively free from some systematics affecting absolute abundances measurements, such as effects associated with non-local thermal equilibrium, 3-dimensional effects in stellar atmospheres, dust depletion and ionisation in diffuse gas. However, the main problem is that one isotope is often much more abundant than the other(s) and their transitions are usually blended, making isotopic abundance ratios difficult to measure.

Due to the large abundance of carbon and the relatively large energy splitting between certain atomic transitions of {\ctwo} and {\cthree}, {\cratio} is one of the easier isotopic abundance ratios to measure. 
{\ctwo} is produced  by helium burning inside massive and short-lived stars as a primary element, whilst {\cthree} is produced within intermediate and low-mass stars by the CNO cycle as a secondary element. 
As a result, {\cratio} is expected to be large ($\gtrsim 1000$) in the early stages of chemical evolution and then decrease with time \citep{Prantzos1996A&A...309..760P,Weischer2010ARNPS..60..381W,kobayashi2011MNRAS.414.3231K}.
Chemical evolution models for the Galaxy predict a slower build up of secondary elements with time (compared to primary elements), and hence a fast decrease in {\cratio} with increasing metallicity \citep{Romano2003MNRAS.342..185R,Kobayashi2020ApJ...900..179K}. 
However, very recent measurements made in stars with $\abundance{Fe}{H}\approx-4$ showed a much lower {\cratio} than predicted \citep{Molaro2023A&A...679A..72M}.  All such stars have measured $\cratiomath<100$, implying a primary production of {\cthree} at low metallicities that conflicts with the predictions \citep{Kobayashi2020ApJ...900..179K}. In particular, several dwarf stars inexplicably have $\cratiomath\lesssim5$ \citep{Molaro2023A&A...679A..72M}. Such low values are close to the theoretical lower limit of $\cratiomath=4$ imposed by the equilibrium value in the CNO bi-cycle \citep{Caughlan1965}.
These new observations suggest that further refinements to chemical evolution models are necessary. In particular, the evolution from very low {\cratio} values at low metallicities to measured solar values \citep[$\cratiomath=91\pm 1.3$,][]{Goto2003ApJ...598.1038G,Ayres2013ApJ...765...46A} remains unexplained.  

Unfortunately, there are only a few {\cratio} measurements in the range $-3 < \abundance{Fe}{H} < -0.2$ \citep{Kobayashi2020ApJ...900..179K}.
Gas at high redshift can provide important supplementary information in a metallicity regime for which there is no information in the Galaxy.
Few measurements or limits exist at cosmological distances (7 measurements are tabulated in \citealt{Romano2022A&ARv..30....7R} with $z_{abs} \ge 0.7$, with one further more recent measurement from \citealt{Welsh_2020MNRAS.494.1411W}). Rotational transitions of molecular gas were used to measure isotopic ratios of several elements, including C, \citep[e.g.][]{Muller2006,Henkel2010A&A...516A.111H,Henkel2014A&A...565A...3H,Wallstrom2016A&A...595A..96W,Noterdaeme2017A&A...597A..82N}. Atomic transitions associated with damped Lyman-$\alpha$ (DLA) absorption systems have also been used to constrain {\cratio} \citep[e.g.][] 
{Levshakov2006A&A...447L..21L,Carswell2011,Welsh_2020MNRAS.494.1411W}. 

Here, we report on {\cratio} in the DLA at $\zabs=1.776$ towards the quasar QSO B1331$+$170, also known as QSO J1333+1649 (R.A.\,13h\,33m\,35.78s, Dec $+$\ang{16;49;04.014}, J2000). This quasar ($V=16.6\,{\rm mag}$, $\zem=2.08895$) was first spectroscopically observed by \citet{Baldwin1973ApJ...185..739B} and the DLA was first characterised by \citet{Strittmatter1973ApJ...183..767S} and  \citet{Carswell1975ApJ...196..351C}. \citet{Meyer1986} used the neutral carbon transitions in this DLA to make the first measurements of the cosmic microwave background radiation temperature, $T_{\rm CMB}$, at high redshift, with later refinements by \citet{Songaila1994}, \citet{Cui2005}, and \citet{Carswell2011}. \citet{Carswell2011} was the first to constrain {\cratio} in this system, finding it must be >5 (at $2\sigma$ confidence limit, CL). Here we present a new {\cratio} constraint derived from new high spectral resolution, high signal-to-noise ratio (S/N), observations with the Echelle SPectrograph for Rocky Exoplanets and Stable Spectroscopic Observations \citep[ESPRESSO, ][]{Pepe2021A&A...645A..96P} on the ESO Very Large Telescope (VLT). A $T_{\rm CMB}$ measurement from the same observations as are used in this paper will be reported in a separate paper.

\section{Data}

\subsection{ESPRESSO observations}
ESPRESSO is an optical (\SIrange{380}{790}{\nano\meter}) echelle spectrograph mounted on the VLT, that can be fed by the light of any one of VLT's four Unit Telescopes. Its high spectral resolution, ($R=\lambda/\Delta\lambda=\SI{140000}{}$, where $\Delta\lambda$ is the full-width at half-maximum of the resolution element), ensures that contributions to the absorption profile shapes from {\ctwo} and {\cthree} can be sufficiently resolved, and its wavelength calibration allows individual exposures to be combined with precision of several \qty{}{\meter\per\second} \citep{Schmidt2021A&A...646A.144S}. ESPRESSO's optomechanical design is intended to ensure that the recorded spectrum does not depend on seeing conditions and that small guiding errors do not cause spectral shapes to vary, thus potentially emulating the presence of {\cthree}. However, ESPRESSO's wavelength coverage provides access only to two \atom{C}{i} transitions ($\lambda1560$ and $\lambda1657$, where the wavelengths are in units {\AA}) of the five known in this DLA (others being $\lambda1277$, $\lambda1280$, and $\lambda1328$). 

We used ESPRESSO in its `singleHR' mode, with one of its two fibres (A) positioned on the object and the second fibre (B) on an empty position \ang{;;7} away for better sky subtraction. Observations were spread across four years, with 12 observations taken in 2019, 11 observations in 2021, and 9 observations in 2023, for a total exposure time of \SI{41.1}{\hour}. The observing log is given in \tabref{tab:observations}, where we have grouped the observations into three epochs (I-III) for easier referencing later in the text. Note that the pixel binning used during detector read-out was $2\times1$ (X$\times$Y, spatial $\times$ spectral direction) in epoch I and $4\times2$ in epochs II and III. 

\begin{table}
    \centering
    \caption{A summary of observations used in this work. The columns provide information on the observing time, detector pixel binning mode (spatial $\times$ spectral direction), exposure time in seconds, airmass at the end of the observations, and VLT Unit Telescope (UT) used, in that order.}
    \label{tab:observations}
    \begin{tabular}{ccccc}
    \hline
    Observing time          & Binning & $T_{\rm exp}$ & Airmass & Telescope\\
    (UTC)                    & (X$\times$Y)  &    (s)    &    &    \\
    \hline
    Epoch I                 &               &         &      &   \\
    2019-04-27T01:30:59.145 & $2\times1$  & 5400 & 1.38 & UT3 \\ 
    2019-04-27T03:02:07.350 & $2\times1$  & 5400 & 1.35 & UT3 \\ 
    2019-04-28T03:47:39.546 & $2\times1$  & 3802 & 1.38 & UT3 \\ 
    2019-04-29T01:15:53.547 & $2\times1$  & 5400 & 1.39 & UT3 \\ 
    2019-04-29T02:46:56.186 & $2\times1$  & 6900 & 1.38 & UT3 \\ 
    2019-05-02T01:09:26.902 & $2\times1$  & 6800 & 1.35 & UT3 \\ 
    2019-05-02T03:03:55.049 & $2\times1$  & 6100 & 1.41 & UT3 \\ 
    2019-05-03T01:01:56.418 & $2\times1$  & 6800 & 1.35 & UT3 \\ 
    2019-05-03T02:56:24.560 & $2\times1$  & 6200 & 1.40 & UT3 \\ 
    2019-05-04T00:54:24.533 & $2\times1$  & 6800 & 1.36 & UT3 \\ 
    2019-05-04T02:48:52.709 & $2\times1$  & 6800 & 1.42 & UT3 \\ 
    2019-05-05T02:54:59.102 & $2\times1$  & 6600 & 1.43 & UT3 \\
    Epoch II &               &     &      &\\
    2021-03-08T07:08:10.623 & $4\times2$  & 3900 & 1.39 & UT3 \\ 
    2021-03-08T08:13:54.843 & $4\times2$  & 3900 & 1.61 & UT3 \\ 
    2021-03-10T06:56:55.242 & $4\times2$  & 4380 & 1.40 & UT3 \\ 
    2021-03-10T08:10:33.381 & $4\times2$  & 4380 & 1.69 & UT3 \\ 
    2021-03-18T06:06:23.381 & $4\times2$  & 4100 & 1.36 & UT1 \\ 
    2021-03-18T07:15:23.208 & $4\times2$  & 4100 & 1.54 & UT1 \\ 
    2021-03-18T08:24:21.701 & $4\times2$  & 4100 & 2.02 & UT1 \\ 
    2021-03-19T05:01:59.645 & $4\times2$  & 4100 & 1.33 & UT1 \\ 
    2021-03-19T06:10:58.690 & $4\times2$  & 4100 & 1.37 & UT1 \\ 
    2021-03-19T07:19:58.152 & $4\times2$  & 4100 & 1.58 & UT1 \\ 
    2021-03-19T08:28:59.043 & $4\times2$  & 3133 & 1.95 & UT1 \\ 
    Epoch III &               &    &      &\\
    2023-02-20T07:12:08.210 & $4\times2$  & 3400 & 1.33 & UT3 \\ 
    2023-02-20T08:15:22.487 & $4\times2$  & 3400 & 1.38 & UT3 \\ 
    2023-02-21T07:54:52.027 & $4\times2$  & 3400 & 1.35 & UT1 \\ 
    2023-02-22T06:47:31.746 & $4\times2$  & 3400 & 1.34 & UT1 \\ 
    2023-02-22T07:48:27.689 & $4\times2$  & 3440 & 1.35 & UT1 \\ 
    2023-02-23T07:01:45.443 & $4\times2$  & 3400 & 1.33 & UT2 \\ 
    2023-02-24T06:44:28.982 & $4\times2$  & 3440 & 1.34 & UT1 \\ 
    2023-02-25T06:06:12.773 & $4\times2$  & 3400 & 1.37 & UT1 \\ 
    2023-02-25T07:06:26.681 & $4\times2$  & 3400 & 1.33 & UT1 \\ 
    \hline
\end{tabular}
\end{table}

\subsection{Data reduction, wavelength calibration, and sky subtraction}
Observations were processed using ESPRESSO Data Reduction Software (DRS) version 3.0.0 \citep{ESPRESSO_DRS3.0.0}, adopting default parameters for all recipes except for the sky subtraction, which was smoothed instead of pixel-by-pixel. The DRS removes detector effects and cosmic rays, and extracts the scientific and the sky spectra from the raw frames using a modified version of the optimal extraction algorithm \citep{Horne1986PASP...98..609H,Robertson_1986PASP...98.1220R}, as described in \citet{Zechmeister_2014A&A...561A..59Z}. The extracted spectra are then corrected for the instrument pixel-to-pixel variations before being wavelength calibrated. ESPRESSO provides two wavelength calibration sources: a Laser Frequency Comb (LFC) and ThAr arc lamp imaged simultaneously with a Fabry-P{\'e}rot etalon. For the majority of our observations, the LFC was unavailable or did not cover the two \atom{C}{i} transitions targeted for this analysis so ThAr + Fabry-P{\'e}rot etalon frames were used to calibrate the spectra presented here. In the final step, a 100-pixel wide sliding average of the sky spectrum is subtracted from the science spectrum.

\subsection{Spectral combination and continuum estimation}
\upl~version 1.05 \citep{Murphy2018_uves_popler_zndo...1297190M,Murphy_2019MNRAS.482.3458M} was used to produce the combined spectrum from reduced observations. \upl~defines a new (log-linear) wavelength grid onto which it redisperses the individual spectra and calculates the inverse-variance weighted average of the values falling within each pixel of the new wavelength grid. Observations taken within a single epoch were combined into a single spectrum, with a wavelength increment of \kmps{0.4} (\kmps{0.8}) for observations taken with $2\times1$ ($4\times2$) detector binning. In producing the combined spectrum, \upl~leverages the availability of the large number of independent flux estimates in each pixel to remove outlying flux values and clip any remaining cosmic rays. \upl~also produces a continuum estimate by fitting a sixth order {\v C}eby{\v s}{\"e}v polynomial over \kmps{2500} wide chunks of the combined spectrum. 

\begin{table}
    \centering
    \caption{S/N per pixel in the continuum for co-added observations. The first column identifies the observation epoch (or their combination), whilst the second column gives the pixel size (in \kmps{}). The last two columns give the S/N values for the two \atom{C}{i} transitions studied here. Note that the values listed for epochs I-III do not simply add in quadrature to produce the combined epoch I+II+III value since spectral dispersions differ.}\label{tab:s/n}
    \begin{tabular}{ccccc}
    \hline
     \multirow{2}{*}{Epoch}  & Pixel size & \multicolumn{2}{c}{Transition}\\
                         & (\kmps{}) & $\lambda1560$ & $\lambda1657$ \\
    \hline
      I   & 0.4 & 32 & 37 \\
      II  & 0.8 & 38 & 41 \\
      III & 0.8 & 38 & 40 \\
     I + I + III  & 0.8 & 70 & 78 \\
     \hline
    \end{tabular}

\end{table}

\subsection{Removing contaminants}\label{sec:contaminants}

\subsubsection{Blends with sky lines and systems at other redshifts}\label{sec:blends}

No sky lines were found to blend with \atom{C}{i} transitions at $\zabs=1.776$. \citet{Carswell2011} reported several transitions at other redshifts in their proximity. \atom{Al}{iii} $\lambda1854$ at $\zabs=1.329$ coincides with the expected positions of \atom{C}{i*} transitions in the proximity of $\lambda1560$, and \atom{Fe}{i} $\lambda2484$ at $\zabs=0.7444$ is nearby. To eliminate any impact from these blends, we set the spectral error array in the wavelength ranges $\SI{4331.4575}{\Angstrom}\leq\lambda \leq \SI{4331.8086}{\Angstrom}$ (\atom{Al}{iii}) and $\SI{4333.5781}{\Angstrom}\leq\lambda \leq \SI{4333.7446}{\Angstrom}$ (\atom{Fe}{i}) to 1000, effectively removing these wavelength ranges from consideration. Our ESPRESSO data does not provide evidence for the existence of the previously reported \citep{Carswell2011} blend between \atom{C}{i} $\lambda1657$ at $\zabs=1.776$ and \atom{C}{iv} $\lambda1550$ at $\zabs=1.9662$ .

\subsubsection{Flux artefacts}\label{sec:artefacts}
Upon combining epoch I spectra, we discovered the unexplained presence of numerous weak but statistically significant flux artefacts in the spectrum. Their presence was first reported in the ESPRESSO spectrum of the quasar HE0515$-$4414 \citep{Murphy2022A&A...658A.123M}. The artefacts appear as weak (a few per cent), positive or negative departures from the average of the surrounding flux values. A typical artefact spans several \kmps{}, appearing at the same (or very similar) positions in the spectrum in each exposure such that traditional outlier rejection methods and spectral combination procedures (as implemented in \upl ) resulted in the increased statistical significance of the artefacts instead of their removal. 

Comparing the three epoch spectra with each other (and other quasar spectra we obtained), we saw that the positions of the artefacts in each epoch spectrum correlated with Earth's motion with respect to the solar system barycentre, indicating an origin that is fixed in the laboratory rest-frame \citep[see figure 2 of][]{Pasquini_2024arXiv240514955P}. The artefacts' origin were tracked to clusters of `warm' detector pixels, that is pixels which erroneously trap an excess of electrons, but which evade detection in hour-long dark exposures in the default DRS settings. 
We identified two warm pixel patches in the proximity of transitions of interest: one in between \qty{4330.93}{\Angstrom} and \qty{4331.14}{\Angstrom}, and another in between \qty{4600.94}{\Angstrom} and \qty{4601.08}{\Angstrom} (wavelengths in the Earth laboratory rest-frame). These wavelength ranges were appropriately shifted to the solar system barycentre and masked in all 32 individual spectra, which were then combined as previously described. The final S/N is reported in \tabref{tab:s/n}.

\section{AI-VPFIT models}\label{sec:models}

This work leverages the recently developed spectral analysis tool \aivpfit~to perform a more comprehensive exploration of the uncertainties \citep{Lee2021aivpfit,Lee2021nonunique,Webb2021,Webb2022,Lee2023}, which expands upon ideas first introduced in \texttt{GVPFIT} \citep{Bainbridge_2017Univ....3...34B,Bainbridge_2017MNRAS.468.1639B}. Previous to this work, limits on {\cthree} abundances were set by examining how $\chi^2$ changes after a specific quantity of {\cthree} has been added into a model of the absorption system containing {\ctwo} only \citep{Levshakov2006A&A...447L..21L}, or by checking how much {\cthree} can be accommodated by the data through simulations of spectra with the same characteristics as the observations \citep{Carswell2011,Welsh_2020MNRAS.494.1411W}.

\begin{figure*}
    \centering
    \includegraphics[trim=0pt 18.8cm 0cm 0cm, clip, width=\textwidth]{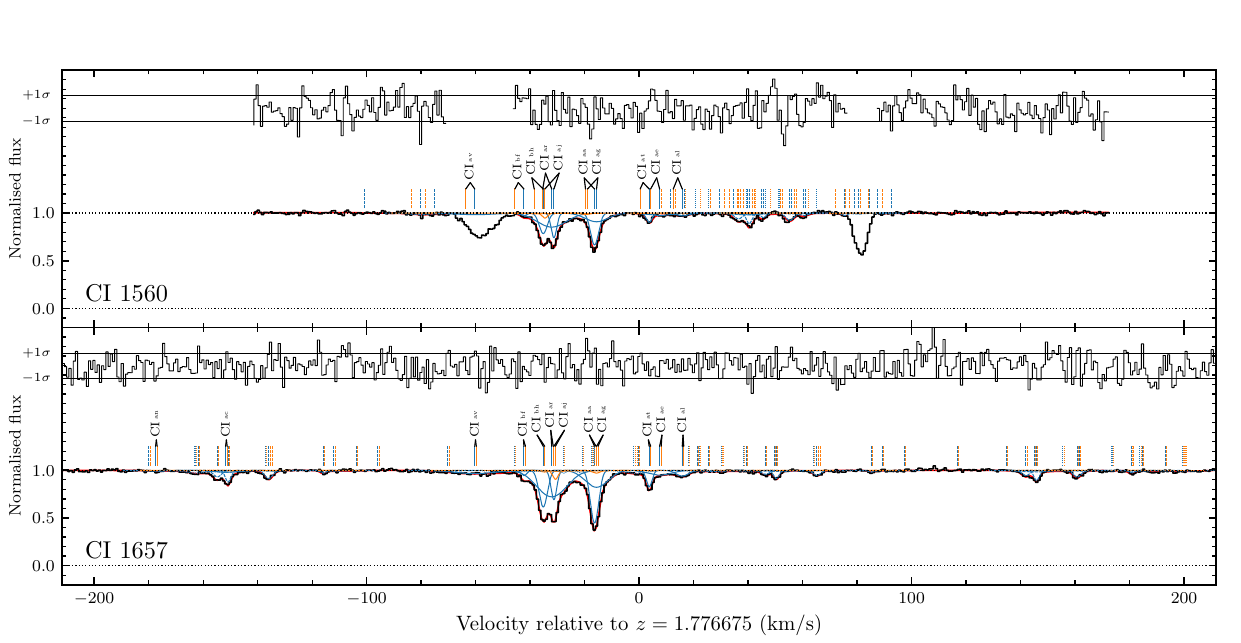}
    \caption{ESPRESSO spectrum of QSO B1331$+$170 in the wavelength region of the two \atom{C}{i} transitions studied here (indicated in the bottom left corner of each panel). The black histogram is the summed spectrum from all three epochs with \kmps{0.8} wide pixels (\figref{fig:model_cratio28p14_separated} shows the individual epoch spectra). The solid red line going through the spectrum shows a 12-component \aivpfit~model with an assumed $\cratiomath=28.14$, close to the final value derived in \secref{sec:analysis}. The model has $\chi^2_\nu=0.6362$. Tick marks above the data show the positions of individual components in the model, with blue and orange ticks showing {\ctwo} and \cthree, respectively. Continuous solid tick marks show ground state transitions (which are the only ones labelled), whereas dashed and dotted ticks show the fine-structure transitions (\atom{C}{i*} and \atom{C}{i**}, respectively). The thin black histogram above the spectrum shows the normalised residuals $({\rm data-model)/error}$. Missing residuals seen in $\lambda1560$ transition indicate spectral regions contaminated by transitions at other redshifts (\atom{Al}{iii} $\lambda1854$ at $\zabs=1.329$ is at \kmps{-60} and \atom{Fe}{i} $\lambda2484$ at $\zabs=0.7444$ is at \kmps{+80}), which were not modelled. Thin horizontal black lines indicate $\pm1\sigma$ ranges. 
    Labels for all transitions are provided in another figure included as supplementary material. 
    }
    \label{fig:spectrum}
\end{figure*}

Given an observed spectrum, {\aivpfit} uses artificial intelligence methods to produce a model for the data, relying on genetic algorithms in combination with a spectroscopic information criterion \citep[SpIC, ][]{Webb2021} to guide the modelling process. SpIC penalises the retention of model parameters fitting weak features more heavily than it penalises parameters fitting strong features. The validity of this modelling system, in terms of obtaining unbiased parameter estimates, has been examined in some detail using simulated spectra by \citet{Lee2021aivpfit, Webb2022}. The simulated spectra were fitted using {\aivpfit} to compare input and estimated parameter values, finding that {\aivpfit} provides robust and unbiased estimates. This is as expected, since \texttt{VPFIT} \citep{Carswell_2014ascl.soft08015C} is embedded within {\aivpfit}, the former being an extensively tested system. Further, spectral models constructed using SpIC retrieved parameter values more accurately than those guided by the corrected Akaike Information Criterion \citep{Akaike1974ITAC...19..716A,Hurvich1989} or the Bayesian Information Criterion \citep{Bozdogan1987}.

Starting from a single-component model in the first generation, {\aivpfit} increases the number of absorbing components in each generation by trying out a user-specified number of trials for additional components (8 in our case) and choosing the one that minimises SpIC. Positions at which the new components are tried is fully determined by a (user specified) seed for the random number generator, making the whole process fully reproducible. This process ends when a set number of attempts fail to further minimise SpIC (30 in our case). {\aivpfit} then performs small adjustments to the model by including modifications to the continuum estimate and tests whether all of the components are necessary (on the basis of SpIC). {\aivpfit} is also capable of identifying blends with unidentified species (interlopers). Having removed known contaminants in \secref{sec:contaminants}, the option of using interlopers was turned off to remove the possibility of fitting real {\cthree} absorption in this way. Further details on {\aivpfit} procedures are given in \citet{Lee2021aivpfit}.

{\aivpfit} was used to produce three hundred independent spectral models to determine $\chi^2$ as a function of assumed {\cthree} abundance in the spectrum and to take into account uncertainties associated with slightly different kinematic structures caused by model non-uniqueness. This was not done before since interactive analysis by a human modeller is time-consuming. {\aivpfit} also allows for examining any biases that may be present by first constructing a model containing {\ctwo} only and to explore the impact of model non-uniqueness on the final result \citep{Lee2021nonunique,Webb2022}.  

\subsection{Resolution and instrumental profile}\label{sec:instrumental_profile}
Theoretical absorption system models are convolved with the assumed shape of ESPRESSO's instrumental profile (IP) before calculating the descent direction in model parameter space. Empirical ESPRESSO IP models were not available at the time, so the IP was approximated to be Gaussian in shape. The resolution for the Gaussian profiles was determined from LFC calibration frames, again available in epoch III, with an extended wavelength range that covers also the two \atom{C}{i} transitions of interest. The full-width at half-maximum (FWHM) was determined in all locations at which the two \atom{C}{i} transitions appear. The final FWHM values used inside {\aivpfit} were their average values: $v_{\rm FWHM}(\lambda1560) = \kmps{2.362}$ and $v_{\rm FWHM}(\lambda1657) = \kmps{2.341}$. Our results are insensitive to small uncertainties on the FWHM values used. For one of the models produced, we explored how best fit parameters change when the FWHM is changed by $\pm1\%$ (based on typical variations seen in LFC calibration frames). We found that model parameter values changed by $\ll1\sigma$, where $\sigma$ here denotes their uncertainties derived from the covariance matrix at best fit.

\subsection{Setting up the calculations} \label{sec:setup}
The calculations were set up to step through $\mathcal{R}\equiv\invcratiomath$ to avoid mathematical infinities associated with low {\cthree} abundances. All {\aivpfit} calculations were set up in exactly the same way, except for the assumed $\mathcal{R}$, which varied from 0.002 to 0.25 in 100 logarithmic steps (sampling was denser at low $\mathcal{R}$). The upper limit on $\mathcal{R}$ is set by the CNO bi-cycle \citep{Caughlan1965}. Three different seeds for the random number generator were used for each $\mathcal{R}$ value, for a total of 300 models. 

Information on the ratio was provided to {\aivpfit} through the `atom.dat' file, containing laboratory wavelengths, oscillator strengths, and natural damping constants ($\Gamma$) for the transitions (see \tabref{tab:atomic_data}). This file was modified in two ways. First, all ground state atomic transitions originating either in {\ctwo} or in {\cthree} were labelled simply as C, such that inserting a C component would automatically result in absorption from both isotopes. The same was also done for \atom{C}{i*} and \atom{C}{i**}, the fine-structure states. Secondly, all oscillator strengths $f$ for the relevant transitions were modified according to the equations: 
\begin{align}
    f^{\rm modified}(^{12}{\rm C}) &= x f^{\rm lab},  \\
    f^{\rm modified}(^{13}{\rm C}) &= (1-x) f^{\rm lab}  ,
\end{align}
where $x = {\rm ^{12}C/C} = 1/(1+\mathcal{R})$ and the superscript \emph{lab} refers to the measured laboratory value (see \tabref{tab:atomic_data}). 

Provided with the modified `atom.dat', {\aivpfit} is able to produce a model for the observations with a pre-determined $\mathcal{R}$. In producing the models, the three epoch spectra were kept separate, and all C species were required to be present in all components with their redshifts tied together. 
Column densities of the tied \atom{C}{i}, \atom{C}{i*}, and \atom{C}{i**} were independent free parameters. A trivial constraint (for numerical convenience) was imposed such that the minimum column density $N$ allowed for any species was $\log (N / \qty{}{\per\centi\meter\squared})= 7.99$. This value is well below any detection threshold in our data and hence has no impact on the results obtained. {\ctwo} and {\cthree} were assumed to have the same line widths, i.e.\ fully turbulent broadening was imposed. We return to this approximation in \secref{sec:summary}. For similar reasons of numerical convenience, the smallest allowed Doppler $b$-parameter was \kmps{0.03}, $\lesssim10\%$ of the pixel size. None of our {\aivpfit} models contain components with $b$-parameters near this limit (the smallest was three times larger).

\begin{figure}
 \includegraphics[width=\columnwidth]{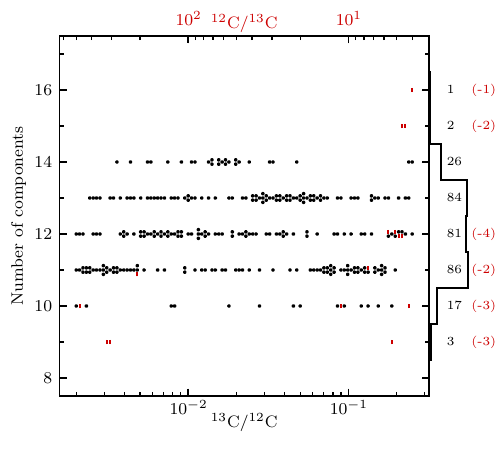}
 \caption{Number of C components, $N_{C}$, for the 300 \aivpfit~models with a histogram. Three models are produced at each $\mathcal{R}$, so multiple models with the same $N_C$ are vertically offset from each other for clarity and visual effect. The total number of models in each group is indicated to the right, where the red numbers in brackets state how many models were removed from the analysis (see text). Removed models are shown as short vertical red lines in the main panel. Ticks in the bottom with black labels show the \invcratio, and ticks on the top with red labels show \cratio.}
 \label{fig:components}
\end{figure}

\begin{table}
    \centering
    \caption{Atomic data for transitions used in this work. The columns give the atomic species, the atomic mass, the laboratory measured wavelength, the laboratory oscillator strength, and the natural damping constant $\Gamma$, in that order. AMU stands for atomic mass unit. Empty fields should be interpreted as repeated entries from one line above. All oscillator strengths are from \citet{Goldbach_1987A&A...181..203G} and all $\Gamma$ values are from \citet{Li_2000JPhB...33.5593L}. Wavelengths for \atom{C}{i} $\lambda1560$ are from \citet{Haridass_1994ApJ...420..433H} and for $\lambda1657$ are from \citet{Lai_2020Atoms...8...62L}. 
    \label{tab:atomic_data}}
    
    \begin{tabular}{lS[table-format=3.1]S[table-format=4.4]S[table-format=3.6]S[table-format=6.2]}
        \hline
         {Species}       & {Atomic mass} & {Wavelength} & {$f^{\rm lab}$} & $\Gamma$  \\
                         & {(AMU)} & {(\AA)}        &                  & {(\qty{}{\per\second})}  \\
         \hline 
         \atom{C}{i}    & 12.0  & 1560.3092 & 0.077400  &  \qty{1.27e+8}{}   \\
                        & 13.0  & 1560.2920 & &  \\ 
         \atom{C}{i**}  & 12.0  & 1560.6823 & 0.058100  &  \qty{1.27e+8}{}   \\
                        & 13.0  & 1560.6650 & &   \\
         \atom{C}{i**}  & 12.0  & 1560.7091 & 0.019300  &  \qty{1.27e+8}{}   \\
                        & 13.0  & 1560.6920 & &   \\
         \atom{C}{i*}   & 12.0  & 1561.3402 & 0.011600  &  \qty{1.27e+8}{}   \\
                        & 13.0  & 1561.3220 & &   \\
         \atom{C}{i*}   & 12.0  & 1561.3667 & 0.000772  &  \qty{1.27e+8}{}   \\
                        & 13.0  & 1561.3500 & &   \\
         \atom{C}{i*}   & 12.0  & 1561.4385 & 0.064900  &  \qty{1.27e+8}{}   \\
                        & 13.0  & 1561.4240 & &   \\
         \atom{C}{i**}  & 12.0  & 1656.2667 & 0.062100  &  \qty{3.61e+8}{}   \\
                        & 13.0  & 1656.2695 & &   \\
         \atom{C}{i}    & 12.0  & 1656.9277 & 0.149000  &  \qty{3.60e+8}{}   \\
                        & 13.0  & 1656.9308 & &   \\
         \atom{C}{i*}   & 12.0  & 1657.0077 & 0.111000  &  \qty{3.61e+8}{}   \\
                        & 13.0  & 1657.0104 & &   \\
         \atom{C}{i**}  & 12.0  & 1657.3788 & 0.037100  &  \qty{3.60e+8}{}   \\
                        & 13.0  & 1657.3812 & &   \\
         \atom{C}{i**}  & 12.0  & 1657.9068 & 0.049400  &  \qty{3.60e+8}{}   \\
                        & 13.0  & 1657.9096 & &  \\
         \atom{C}{i*}   & 12.0  & 1658.1206 & 0.037100  &  \qty{3.61e+8}{}  \\
                        & 13.0  & 1658.1234 &    &    \\
         \hline
        \end{tabular}
        
\end{table}

\subsection{Quality check of the models}\label{sec:quality_checks}
\begin{figure}
 \includegraphics[width=\columnwidth]{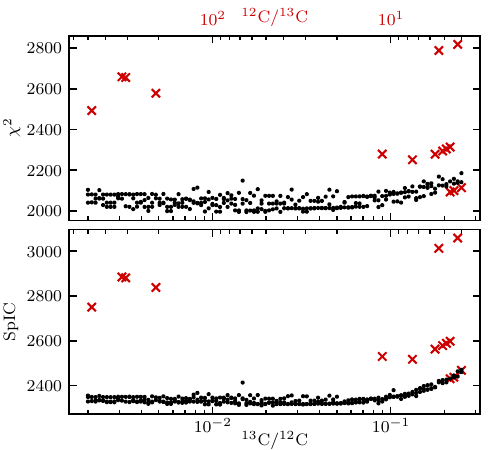}
 \caption{$\chi^2$ (top panel) and SpIC (bottom panel) for the 300 \aivpfit~models, as a function of {\invcratio} (bottom ticks and black labels) or {\cratio} (top ticks and red labels) on a logarithmic scale. Black dots are the 285 points used in the analysis and 15 models plotted as red crosses have been discarded (see text). }
 \label{fig:chisq+spic}
\end{figure}

\figref{fig:spectrum} shows one of the {\aivpfit} models with 12 velocity components. \figref{fig:components} shows the number of components ($N_{C}$) for all models as a function of $\mathcal{R}$, with a corresponding histogram. No models have been produced with fewer than 9 or more than 16 components, with the median number of components being 12. Interestingly, while models with $N_{C}=\SIrange{9}{14}{}$ were produced over a wide range of $\mathcal{R}$ values, the two models with $N_{C}=15$ and the one model with $N_{C}=16$ are associated only with high $\mathcal{R}$. Examining the latter three models revealed them not to fit well the strong feature at \kmps{-17} in \atom{C}{i} $\lambda1657$ (the residuals systematically deviate from zero), so they were discarded\footnote{The ($\mathcal{R},\chi^2$) values of these models are (0.215973,2093.7281), (0.226768,2100.352), and (0.250001,2114.7071), i.e.\ they are marked by the three crosses sitting below the black points in the top panel of \figref{fig:chisq+spic}.}. Similarly, all three models with $N_{C}=9$ were found to erroneously fit the ``double'' absorption feature (located between \kmps{-30} and \kmps{-36} in \figref{fig:spectrum}) using a single component, and were also discarded. Visual examination of the remaining models did not show obvious problems with fitting the observations. We show several other models in Appendix \ref{sec:example_models}. Plots of all models are available as online supplementary material.  

Also interesting is the apparent preference towards models with a specific $N_{C}$ for a given $\mathcal{R}$. For example, fewer components are generally required when $\mathcal{R}$ is assumed to be at either extreme of the examined range. Models requiring the largest number of components (14) prefer $\mathcal{R}$ between 0.01 and 0.02, 
and do not spread in $\mathcal{R}$ like all the others (ignoring the 2 points at high $\mathcal{R}$). $N_{C}=13$ models seem to preferentially appear between $\mathcal{R}\approx0.02$ and 0.08 and models with $N_{C}=12$ do not appear to cluster significantly at any $\mathcal{R}$ value. $N_{C}=11$ models are preferred for two $\mathcal{R}$ ranges: between 0.07 and 0.2, and below 0.005. This split is suggestive of different velocity structures and hence model non-uniqueness. On the other hand, $N_{C}=10$ appear at all $\mathcal{R}$ values but are not very common. 

The top panel of \figref{fig:chisq+spic} shows the $\chi^2$ for all 300 models, calculated as:
\begin{equation}\label{eq:chisq}
    \chi^2= \sum_{i=1}^N \left( \frac{  F_i - \mathcal{F}_i}{\sigma_i } \right)^2.
\end{equation}
Above, $F_i$ is the observed flux in the $i$\textsuperscript{th} pixel, $\mathcal{F}_i$ is its model predicted value, and $\sigma_i$ is the observed flux error estimate \citep[in our case, the root-mean-square array produced by \texttt{UVES\_popler} was used in place of the spectral variance array produced by the pipeline, for details see][]{Murphy_2019MNRAS.482.3458M}. The bottom panel of the same Figure shows ${\rm SpIC = \chi^2 + } $ a penalty term that is a function of model parameters \citep{Webb2021}, with the penalty term being different for each model. We note the smaller scatter in SpIC compared to $\chi^2$, a general trend seen before in \citet{Lee2021nonunique}. See further comments in Section \ref{sec:pdf+cdf_spic}. The smallest $\chi^2$ and SpIC values appear between $\mathcal{R}=0.01$ and 0.04, corresponding to $\cratiomath$ between 25 and 100.

Twelve models show anomalously large $\chi^2$ values (3, 3, 2, 4 models with $N_{C}=9,10,11,12$, respectively, some of which were previously identified as problematic). These models were also rejected. The reduced $\chi^2_\nu \equiv \chi^2 / \nu$ ($\nu$ being the number of degrees of freedom) of the 285 models ranges between 0.621 and 0.678, with 0.638 being the median value. These values are clearly slightly below the expectation of around unity. The reason is well understood: re-dispersion of spectra and their subsequent combination create adjacent pixel correlations in the combined spectrum. Parameter error estimates take this into account, as described in \citet{web:VPFIT}. Since the spectra from three epochs were kept as separate entities during the modelling, we spot-checked for consistency (goodness of fit) for each epoch. In cases examined, the $\chi^2$ values for corresponding regions were fully consistent, i.e. the overall model fitted each epoch equally well.

\subsection{Inter-model variation and non-uniqueness}\label{sec:nonuniqueness}

\begin{figure*}
    \includegraphics[width=\textwidth]{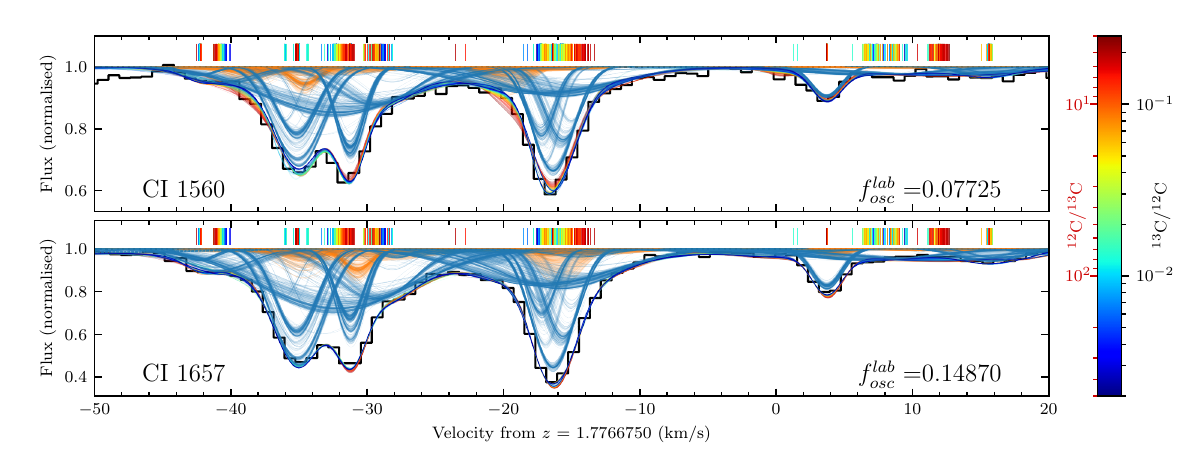}
    \caption{Illustration of model variation and non-uniqueness. The astronomical data are shown as a black histogram (combined epochs I+II+III, \kmps{0.8} wide bins). Individual models are shown as coloured lines, with their colour corresponding to a specific {\cratio} value, as per the colour bar to the right. Coloured tick marks show the positions of {\ctwo} absorption in those models (also as per the colour bar). There are 2542 \atom{C}{i} components shown from all 285 models (more components fall outside the plotted velocity range, but they fit weaker absorption features). Thin blue lines show {\ctwo} absorption and thin orange lines show {\cthree} absorption after convolution with the assumed IP. Not all components seen in the Figure are present in all models (see \secref{sec:nonuniqueness}). Note the different ordinate scales for the two panels.
    }
    \label{fig:nonuniqueness}
\end{figure*}

As described above, a large number of models were computed. We used 100 different settings for {\cratio}, and at each {\cratio} compute 3 independent {\aivpfit} models. At fixed {\cratio}, each of the 3 models is constructed independently, in the sense that trial absorption component placement is random within the fitting range. Model development therefore proceeds differently each time. This emulates, in a rather natural way, the notion of using a large number of independent interactive modellers, and hence avoids bias. The consequence of this is that slight variations from one model to the next are seen in the ensemble of {\aivpfit} models. We illustrate this in \figref{fig:nonuniqueness} by showing both the astronomical data and the 285 surviving models (see \secref{sec:quality_checks}) superimposed. Tick marks show the positions of velocity components for all 285 models, for the dominant section of the absorption system (compare Figures \ref{fig:spectrum} and \ref{fig:nonuniqueness}). The colour coding provides further information, since it shows the {\cratio} for each model. In \figref{fig:nonuniqueness}, one must appreciate that not all components illustrated are present in all models. For example, one {\aivpfit} model might opt to use a single velocity component whilst another might prefer to use two weaker ones. The summed optical depths over all models shown therefore is not intended to represent a good fit to the astronomical data, but is instead intended to illustrate model variation.

There are at least three important features for constraining {\cratio}: at  \kmps{-38}, \kmps{-20}, and \kmps{+4}. {\cratio} is not allowed to be too high, otherwise there will be excess absorption in the blue wing of \atom{C}{i} $\lambda1560$ at \kmps{-38}. A low {\cratio} would result in insufficient absorption at \kmps{-20} in \atom{C}{i} $\lambda1560$, but a high {\cratio} would cause the model to fall below the data. In the \kmps{+4} {\ctwo} feature, some {\cthree} is needed in $\lambda1560$ to fit weak absorption at \kmps{+1} and similarly at \kmps{+5} in $\lambda1657$. Interestingly, the feature at \kmps{+4} is the only one that is consistently fitted by a single component in all 285 models. Determining which of these three features provides the most stringent constraint is not easy. Isolating any particular component to investigate that component's relative importance in constraining the overall {\cratio} is difficult; artificially freezing some parameters whilst allowing others to vary is likely to make error estimates for parameters of interest unreliable. Irrespective, as \figref{fig:nonuniqueness} shows, several velocity components contribute to the overall constraint.

The ``AI-Monte Carlo'' approach used in this work allows us to examine, in detail, model non-uniqueness, which can be seen visually in \figref{fig:nonuniqueness}. It is interesting that the {\aivpfit} models of ESPRESSO data are in broad agreement with previous measurements for the physical characteristics of the gas by \citet{Carswell2011}, although the improvement in spectral resolution offered by ESPRESSO means far more components are resolved. Further details can be found in Appendix \ref{sec:theoretical_models}. In the following Section, we develop methods to simultaneously use the set of model calculations in order to place simple constraints on the {\cratio} averaged over the absorption complex. We also show how, in this particular case, non-uniqueness turns out to create ``groupings'' in {\cratio} space.

\section{Analysis}
\label{sec:analysis}

The likelihood of $\mathcal{R}$ is:
\begin{equation}\label{eq:likelihood}
    \mathcal{L} (\mathcal{R}) = \exp{\left[- \frac{1}{2} \chi^2(\mathcal{R}) \right]}.
\end{equation}

$\chi^2$ minimises ($\mathcal{L}$ maximises) for the physically correct $\mathcal{R}$, and can be approximated by a parabola around that value:
\begin{equation}\label{eq:parabola}
    \chi^2(\mathcal{R}) \approx a(\mathcal{R}-\mathcal{R}_{min})^2 + \chi^2_{min} ,
\end{equation}
where $a$ defines the parabola's openness, and $\mathcal{R}_{min}$ and $\chi^2_{min}$ are the vertex coordinates. The $\chi^2$ values of the 285 models are assumed to scatter normally around Equation \eqref{eq:parabola} prediction with variance $\varsigma^2$. Equation \eqref{eq:likelihood} must be regarded as an approximation because the terms in the $\chi^2$ summation are not independent.

\subsection{Dividing the models into five subsets}\label{subsec:data_division}
\begin{figure}
 \includegraphics[width=\columnwidth]{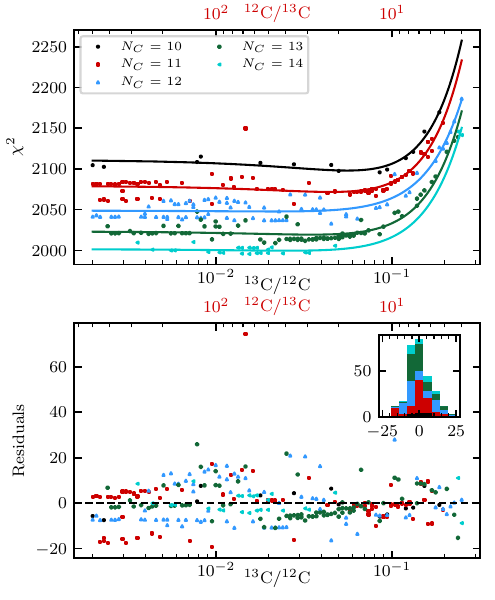}
 \caption{Results of MCMC fitting of Equation \eqref{eq:parabola} to observed $\chi^2$ values, subdivided according to $N_{C}$. Top panel: Coloured symbols are 285 models and the coloured lines are the best fitting parabolas going through them. Parabola parameters are independent from each other. Bottom panel: Residuals (data$-$model) of the fits shown in the top panel. Some apparently correlated structures remain (see text). The inset shows a stacked histogram of the residuals. In both panels, ticks in the bottom with black labels show the \invcratio, and ticks on the top with red labels show {\cratio}.} 
 \label{fig:parabola}
\end{figure}

\figref{fig:parabola} shows there is a clear stratification in $\chi^2$, most of which can be attributed to the number of free model parameters, that is to the number of carbon components in the model ($N_{C}$). There is also a possible bifurcation for models with the same $N_{C}$: models at similar $\mathcal{R}$ appear systematically offset in $\chi^2$ by approximately the same amounts, apparently grouping together. Models with $N_{C}=\qtylist{11;12;13}{}$ (red squares, blue triangles, and green diamonds, respectively) show this effect most clearly. Comparison of models and their parameters did not reveal any clues on the origin of this effect.

$\chi^2$ values were divided into subsets according to their $N_C$, and were henceforth treated independently. The five groups are shown by symbols of different colours in \figref{fig:parabola}. Parabola parameters (and their uncertainties) for each subset were determined using MCMC calculations \citep[implemented in \texttt{numpyro},][]{Phan2019_numpyro} such that the uncertainties can be propagated into constraints on $\mathcal{R}$. Parameters $a$, $\chi^2_{min}$, and $\mathcal{R}_{min}$ were all sampled from uniform prior distributions having the following ranges: (2000,\qty{5000}{}),  (1900,2200), (0, 0.25), in the same order. $\varsigma$ was independently determined for all five subsets, having an exponential distribution prior with a scale of unity. Ten MCMC chains were ran with \SI{7000} warm-up steps and \SI{2500} sample steps. Convergence was assessed using two diagnostics: the split Gelman-Rubin $\hat{R}$ \citep{Rhat} and effective sample size $N_\mathrm{eff}$ \citep{Geyer_introMCMC}. For all parameters of the five parabolas, $\hat{R}=1.00$ and $N_\mathrm{eff}\gg1000$, meaning that convergence was achieved. Marginalised posterior distributions of the fitting parameters and their covariances are provided in Appendix \ref{sec:mcmc}, together with their median values and uncertainties. 

Coloured lines in the top panel of \figref{fig:parabola} show the median MCMC predictions for the data. Grouping by $N_{C}$ explains much of the observed scatter between the points. Residuals (bottom panel of the same Figure) show that some substructure remains. Because absorption system models were all derived from the same spectral data, their $\chi^2$ values are not expected to scatter completely randomly, and the remaining substructure may be irrelevant. Alternatively, there may be an additional  contribution to the scatter, beyond the one attributable to $N_C$, for example model non-uniqueness.

\begin{figure*}
 \includegraphics[width=\textwidth]{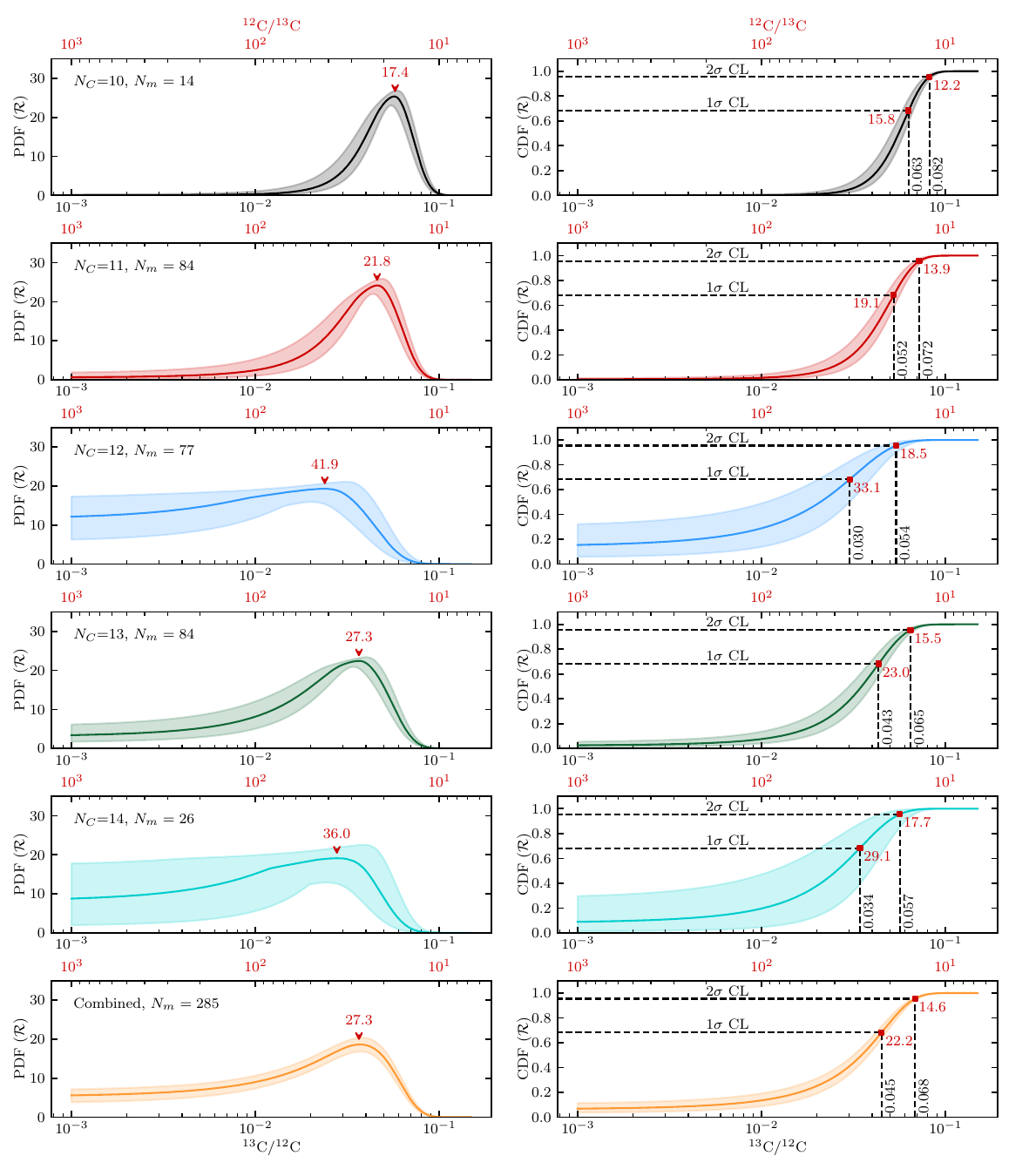}
 \caption{PDF($\mathcal{R}$), left, and CDF($\mathcal{R}$), right, for the five data subsets and the full data set. Each row shows one subset of the data, with the inset specifying $N_{C}$ and the number of models contained in the subset, $N_m$. The combined PDF and CDF (final row) are averages of the five above weighted by $N_m$. In all panels, the thick coloured line is the median over parabola parameters derived by MCMC, and the shaded bands enclose the central 68\% of the possible curves. In the panels showing the PDF, the red arrow with a number on top indicates the most probable {\cratio} (the mode of the distribution). In the panels showing the CDF, the horizontal dashed black lines show the probabilities corresponding to $1\sigma$ and $2\sigma$ confidence limits. Red squares show where those lines intersect the CDF, and the vertical dashed black lines show the corresponding \invcratio, printed in black. The red numbers adjacent to the red squares are their inverse, i.e.\ the {\cratio}. Ticks in the bottom with black labels show the \invcratio, and ticks on the top with red labels show \cratio. }
 \label{fig:pdf_cdf}
\end{figure*}

\subsection{Constraining $\mathcal{R}$ using $\chi^2$}\label{sec:pdf+cdf}

Substituting Equation~\eqref{eq:parabola} into \eqref{eq:likelihood} and normalising, the probability density function (PDF) and the cumulative probability density function (CDF) can be calculated analytically from:
\begin{align}
    {\rm PDF} (\mathcal{R}, \mathcal{R}_{min}, a ) =& \sqrt{\frac{a}{2\pi}} \exp{ \left(-\frac{a}{2}\left(\mathcal{R} - \mathcal{R}_{min}\right) ^2 \right)}, \label{eq:pdf} \\
    {\rm CDF} (\mathcal{R}, \mathcal{R}_{min}, a) =& \frac{1}{2} \left[1 + \erf\left({\sqrt{\frac{a}{2}}(\mathcal{R} - \mathcal{R}_{min}) }\right)\right]. \label{eq:cdf} 
\end{align}

\figref{fig:pdf_cdf} shows the results for the PDF (left column) and the CDF (right column) derived from \qty{25000}{} different parameter combinations obtained from the MCMC calculations in \secref{subsec:data_division}. The first five rows relate to the five data subsets. The last row shows the average of the first five weighted by the number of values in each subset, $N_m$. 

$N_{C}=10$ case provides the most stringent constraints on $\mathcal{R}$, followed by $N_{C}=\qtylist{11;13;14;12}{}$ (in that order). For three subsets ($N_{C}=\qtylist{10;11;13}{}$), PDFs are well localised and indicate a statistically significant measurement. Modes of the five curves fall within $\cratiomath=\qtyrange{17.4}{41.9}{}$, in good agreement with each other (when compared to the width of their PDFs). PDFs of $N_{C}=\qtylist{12;14}{}$ permit the broadest range of solutions, including $\cratiomath>1000$. Plotting a subset of $N_{C}=\qtylist{12;14}{}$ curves (not shown) we identified some which yielded results consistent with the remaining three subsets. Future, improved, data may reveal which ones are correct. Our results here indicate that the probability of $\cratiomath<100$ is \qtylist{99.86;98.43;71.17;92.44;80.44}{\percent} (from $N_C=10$ to 14, respectively). The combined PDF (last row, left column) is well localised with a mode at $\cratiomath=27.3$ and the combined probability that $\cratiomath<100$ is \qty{86.48}{\percent}. Some other statistical properties, including percentiles and lower limits, are summarised in \tabref{tab:cratio_measurements}.

\begin{table}
    \centering
    \caption{{\cratio} measurement summary. The first column indicates data subset ($N_{C}$ or combined). The second column gives the most probable value for {\cratio} for each case, i.e.\ the mode of the PDF. The following three columns provide percentile values for {\cratio}. The ultimate two columns give lower $1\sigma$ (68.27\% probability) and $2\sigma$ (95.45\% probability) confidence limits on \cratio, derived from the CDFs. 
    Note, although we use the $\sigma$ terminology here, it is clear from \figref{fig:pdf_cdf} that the uncertainties are not Gaussian. 
    \label{tab:cratio_measurements}}
    \begin{tabular}{cS[table-format=3.1]S[table-format=3.1]S[table-format=3.1]S[table-format=3.1]S[table-format=3.1]S[table-format=3.1]}
    \hline
    \multirow{2}{*}{Subset} & {\multirow{2}{*}{Mode}} & \multicolumn{3}{c}{Percentiles} & \multicolumn{2}{c}{CL}\\
     &  &    {16th} &     {50th} &      {84th} &   {$1\sigma$}  & {$2\sigma$} \\
    \hline
           10 &   17.4 &  14.1 &  17.9 &  24.7 &           15.8 &           12.2 \\
           11 &   21.8 &  16.4 &  22.3 &  34.9 &           19.1 &           13.9 \\
           12 &   41.9 &  24.8 &  47.8 & 693.5 &           33.1 &           18.5 \\
           13 &   27.3 &  19.1 &  28.5 &  56.5 &           23.0 &           15.5 \\
           14 &   36.0 &  22.8 &  39.0 & 133.5 &           29.1 &           17.7 \\
           Combined &   27.3 &  18.1 &  28.5 &  80.0 &           22.2 &           14.6 \\
    \hline
\end{tabular}

\end{table}

\subsection{Constraining $\mathcal{R}$ using SpIC}\label{sec:pdf+cdf_spic}

In the preceding Section, we used $\chi^2$ as the basis for computing likelihood values. An alternative statistical approach is to instead use an information criterion \citep{Burnham2002}, in this case SpIC. We carried out tests, replicating the procedure used for the $\chi^2$ results shown in Table \ref{tab:cratio_measurements}. The preliminary indication was that, provided sample subdivision is used, the results do not change significantly. If sample subdivision is not used, the tighter grouping for the SpIC measurements (see Figure \ref{fig:chisq+spic}) suggests that the SpIC values yield better constraints on {\cratio}. We also note that Bayesian Model Averaging could also be applied using an information criterion likelihood approach. This would offer the simplification of deriving one numerical result from the entire suite of {\aivpfit} models. However, preliminary calculations suggested a strong weighting towards lowest SpIC models, giving an overly stringent result, which we preferred to avoid in this context.

\section{Discussion}\label{sec:discussion}

\subsection{Comparison with previous measurements}

The result reported here is $\cratiomath=28.5^{+51.5}_{-10.4}$. \citet{Carswell2011} reported a conservative lower limit of $\cratiomath>5$ ($2\sigma$ CL) in this DLA. Looking at their figure 2, {\cthree} was found only in their component N2 (at \kmps{-17} in \figref{fig:spectrum}), even though N2 appears (in their spectrum) significantly weaker than component N1 (at \kmps{-33} in \figref{fig:spectrum}). However, our higher resolution ESPRESSO data revealed that their component N1 is actually a blend of several absorption features. No significant substructure is seen for their component N2 in the ESPRESSO spectrum, which also appears as the strongest feature in the absorption complex. The new ESPRESSO data therefore explains why {\cthree} seemed to arise in the weaker absorption component in the \citet{Carswell2011} analysis. Interestingly, their reported ratio between column densities of {\ctwo} and {\cthree} in N2 equates to $\cratiomath=19^{+20}_{-10}$, in agreement with our findings. Considering that the inferred column densities are close to the linear part of the curve of growth, this agreement is not surprising. 

Only three other {\cthree} quasar absorption measurements currently exist. Two recent values are: \citet{Noterdaeme2017A&A...597A..82N} constrained $\cratiomath>40$ at $\zabs=2.525$; \citet{Welsh_2020MNRAS.494.1411W} constrained $\cratiomath>2.3$ at $\zabs=2.34$. Another seemingly more stringent result is quoted in \citet{Levshakov2006A&A...447L..21L}, who measured {$\cratiomath>80$} ($2\sigma$ CL) at $\zabs=1.15$, although an independent re-analysis of the same data did not support such a high value and instead found {$\cratiomath>22$} \citep{Carswell2011}. Table \ref{tab:cratio_measurements} provides our analogous $2\sigma$ lower limit, 14.6, i.e. all quasar results to date appear consistent.

\subsection{Chemical evolution models}

{\ctwo}  is formed in the triple-$\alpha$ process during hydrostatic helium burning and is a primary product of stellar nucleosynthesis.  The stable$^{13}$C isotope is  produced in the hydrogen-burning shell when the CN cycle converts pre-existing {\ctwo} into $^{13}$C via proton capture followed by $\beta$ decay. 
As a  star evolves off the main sequence,  the outer convective envelope expands
inwards  into the CN-cycle-processed regions producing a  mixing episode which is  called the `first dredge-up', which    lowers the {\cratio} \citep{iben1984PhR...105..329I}. Mixing  also occurs  in the thermal pulses of intermediate-mass stars that become AGB stars. This could lead to a CN-cycle equilibrium ratio of about   $\cratiomath\sim4$. These low values ratios are indeed observed in some red supergiants \citep{lambert1977ApJ...215..597L}.

Due to the secondary nature of {\cthree}, chemical evolution models  predict a monotonic decrease in the isotopic ratio {\cratio} with time \citep{Romano2003MNRAS.342..185R,Kobayashi2020ApJ...900..179K}. This is supported by observations   in   young molecular clouds for which  $\cratiomath\sim 60-70 $, lower than  the  solar ratio of  $\cratiomath = 91\pm 1.3$ \citep{Goto2003ApJ...598.1038G,Ayres2013ApJ...765...46A}. {\cratio} is also below solar value in the Galactic centre \citep{Halfen2017ApJ...845..158H} and measurements in the Galaxy support a gradient with galactocentric distance \citep{yan2023A&A...670A..98Y}. All of the above is consistent with the secondary nature of {\cthree}. For the reasons above, {\cratio}  is predicted to be  higher than  solar at low metallicities \citep{Romano2003MNRAS.342..185R,fenner2005MNRAS.358..468F,Kobayashi2020ApJ...900..179K,Romano2022A&ARv..30....7R}. Unfortunately, only a few observations are available in this regime and the results are controversial. \citet{botelho2020MNRAS.499.2196B}  found   a mild increase in the  range $-0.2< \abundance{Fe}{H}<0$  in a sample of  solar twins, opposite to what is expected if {\cthree} is a secondary element. On the other hand,  \citet{Crossfield2019ApJ...871L...3C} measured $\cratiomath = 296\pm 45$ and $224 \pm 26$ in the two components of the  brown dwarf system GJ 745 at $\abundance{Fe}{H}=-0.48$.  In some unmixed giants {\cratio} is lower than the solar value \citep{spite05} and in the  metal-poor star HD 140283, $\cratiomath = 33^{+12}_{-6}$  \citep{spite2021A&A...652A..97S}. Moreover,  very low values of {\cratio} have been derived for several CEMP-no stars (carbon enhanced metal poor stars with $\abundance{Ba}{F}<0$) with metallicities lower than  $\abundance{Fe}{H}\lesssim$ -4 \citep{Molaro2023A&A...679A..72M}. At very low metallicities the disagreement with the theoretical expectations is striking since values of several thousands are foreseen due to an initial {\cthree} close to zero. Internal production and chemical transfer from a possible massive companion have been  ruled out, so the {\cthree} enhancement must originate from their progenitors.

At very low metallicities,  mixing between the H- and He-burning zones could be   driven by rapid rotation leading to a production   of  {\cthree}   \citep{Meynet2006A,chiappini2008A&A...479L...9C,Limongi2018}.   Significant  quantities of {\cthree}  are expected  from massive, low-metallicity, fast-rotating stars with {\cratio} between 30 and 300 \citep{chiappini2008A&A...479L...9C}. Processes necessary to produce {\cthree}  are similar to those invoked to explain the primary N behaviour at very low metallicities in other DLAs \citep{molaro2003ASPC..304..221M,zafar2014MNRAS.444..744Z}.

No stellar {\cratio} measurements are available at the metallicity of DLA studied here  \citep[$\abundance{Fe}{H}=-1.27$, ][]{Berg2015MNRAS.452.4326B} so no direct comparison is possible. Our robust lower limit ($\cratiomath>14.6$, $2\sigma$ CL) is consistent with both theoretical predictions and low metallicity measurements. On the other hand, the weighted value we obtained ($\cratiomath=28.5$, see \tabref{tab:cratio_measurements}) is consistent with the measurement  in the star HD 140283 at [Fe/H] $\approx$ -2.6, but much lower than that of GJ 745 at $\abundance{Fe}{H}=-0.48$.  If our measured value in the DLA studied here is typical of gas with metallicity of $\abundance{Fe}{H}\approx$ -1.2, it would imply there is a significant and early production of $^{13}$C that must necessarily cease to allow {\cratio} to reach the measured values at $\abundance{Fe}{H}\approx -0.5$.

\section{Summary}\label{sec:summary}
In this work, we have carried out a detailed {\aivpfit} study of the $\zabs=1.776$ DLA towards QSO B1331+170 for a {\cratio} measurement. The spectral data are very high quality and hence expected to substantially tighten previous error bars for this system. However, perhaps unsurprisingly, the higher ESPRESSO spectral resolution detected structure that had previously been unresolved, such that the new measurement constraints were not as anticipated. 

The use of {\aivpfit} permitted us to generate multiple independent models. This means that, for the first time in this context, we have been able to take into account measurement errors associated with model non-uniqueness (\secref{sec:nonuniqueness}). Our weighted final result (\secref{sec:pdf+cdf}) is $\cratiomath=28.5^{+51.5}_{-10.4}$ (see \tabref{tab:cratio_measurements} and \figref{fig:pdf_cdf}), a marginal {\cthree} detection. Additional ESPRESSO or new ANDES \citep{Marconi_2022SPIE12184E..24M,Marconi_2024arXiv240714601M} or G-CLEF \citep{Szentgyorgyi_2018SPIE10702E..1RS} observations may lead to more stringent constraints in the future. A further caveat is that our {\aivpfit} analysis required the {\cratio} to be the same in all components. Spatial {\cratio} variations could be present within the absorption complex, so our result represents the mean value over the system. 

Finally, as mentioned in Section \ref{sec:setup}, we have assumed constant $b$-parameters for both {\ctwo} and {\cthree}. Since the atomic masses are so similar, modelling the system using thermal broadening is impractical, as numerical instabilities would be introduced into the modelling process. If the overall line broadening contains a thermal contribution, the slightly different atomic masses means our assumption is not quite correct. The maximum impact of this approximation equates to a systematic 4\% of the overall $b$-parameter. The best fractional uncertainty of any $b$ in the absorption complex we measured is slightly below this value, which means the constant-$b$ approximation may impose a slight systematic if there is a significant contribution to thermal broadening. However, the direction of this potential small systematic is such that our lower limit on {\cratio} would increase, that is, our quoted final result is perhaps slightly conservative.

\section*{Acknowledgements}
The authors acknowledge the ESPRESSO project team for its effort and dedication in building the ESPRESSO instrument as well as the teams who developed software used for the analysis -- \texttt{numpy} \citep{harris2020array_numpy}, \texttt{scipy} \citep{2020SciPy-NMeth}, \texttt{matplotlib} \citep{Hunter_2007Matplotlib}, \texttt{numpyro} \citep{Phan2019_numpyro}, \texttt{jax} \citep{jax2018github}, and \texttt{jaxopt} \citep{jaxopt_implicit_diff}. 
This work was performed on the OzSTAR national facility at Swinburne University of Technology. The OzSTAR program receives funding in part from the Astronomy National Collaborative Research Infrastructure Strategy (NCRIS) allocation provided by the Australian Government, and from the Victorian Higher Education State Investment Fund (VHESIF) provided by the Victorian Government. 
The INAF authors acknowledge financial support of the Italian Ministry of Education, University, and Research with PRIN 201278X4FL and the "Progetti Premiali" funding scheme. 
PJ is funded by the Austrian Science Fund (FWF): F6811-N36 within the SFB F68 “Tomography Across the Scales”. 
MTM acknowledges the support of the Australian Research Council through Future Fellowship grant FT180100194.
TMS acknowledges the support from the SNF synergia grant CRSII5-193689 (BLUVES). 
This work was financed by Portuguese funds through FCT (Funda\c c\~ao para a Ci\^encia e a Tecnologia) in the framework of the project 2022.04048.PTDC (Phi in the Sky, DOI 10.54499/2022.04048.PTDC). CJM also acknowledges FCT and POCH/FSE (EC) support through Investigador FCT Contract 2021.01214.CEECIND/CP1658/CT0001 (DOI 10.54499/2021.01214.CEECIND/CP1658/CT0001). 
JIGH and ASM acknowledge the financial support from the Spanish Ministry of Science and Innovation (MICINN) project PID2020-117493GB-I00. 
We acknowledge financial support from the Agencia Estatal de Investigaci\'on of the Ministerio de Ciencia e Innovaci\'on MCIN/AEI/10.13039/501100011033 and the ERDF “A way of making Europe” through project PID2021-125627OB-C32, and from the Centre of Excellence “Severo Ochoa” award to the Instituto de Astrofisica de Canarias.
NCS is co-funded by the European Union (ERC, FIERCE, 101052347). Views and opinions expressed are however those of the author(s) only and do not necessarily reflect those of the European Union or the European Research Council. Neither the European Union nor the granting authority can be held responsible for them. This work was supported by FCT through national funds and by FEDER through COMPETE2020 - Programa Operacional Competitividade e Internacionaliza{\c c}{\~a}o by these grants: UIDB/04434/2020; UIDP/04434/2020.
SGS acknowledges the support from FCT through Investigador FCT contract nr.\ CEECIND/00826/2018 and POPH/FSE (EC).

\section*{Data Availability}

Based on observations collected at the European Southern Observatory under ESO programmes 1102.A-0852(C), 106.218R.002, 109.2335.002, and 111.24NH.002 (PI for all programmes is Paolo Molaro) as a part of the ESPRESSO Guaranteed Time Observations. Unprocessed observations are availabile through the ESO Archive. Spectra and {\aivpfit} models produced as a part of this work are provided as supplementary online material.



\bibliographystyle{mnras}
\bibliography{ref} 

\begin{thebibliography}{}
\makeatletter
\relax
\def\mn@urlcharsother{\let\do\@makeother \do\$\do\&\do\#\do\^\do\_\do\%\do\~}
\def\mn@doi{\begingroup\mn@urlcharsother \@ifnextchar [ {\mn@doi@}
  {\mn@doi@[]}}
\def\mn@doi@[#1]#2{\def\@tempa{#1}\ifx\@tempa\@empty \href
  {http://dx.doi.org/#2} {doi:#2}\else \href {http://dx.doi.org/#2} {#1}\fi
  \endgroup}
\def\mn@eprint#1#2{\mn@eprint@#1:#2::\@nil}
\def\mn@eprint@arXiv#1{\href {http://arxiv.org/abs/#1} {{\tt arXiv:#1}}}
\def\mn@eprint@dblp#1{\href {http://dblp.uni-trier.de/rec/bibtex/#1.xml}
  {dblp:#1}}
\def\mn@eprint@#1:#2:#3:#4\@nil{\def\@tempa {#1}\def\@tempb {#2}\def\@tempc
  {#3}\ifx \@tempc \@empty \let \@tempc \@tempb \let \@tempb \@tempa \fi \ifx
  \@tempb \@empty \def\@tempb {arXiv}\fi \@ifundefined
  {mn@eprint@\@tempb}{\@tempb:\@tempc}{\expandafter \expandafter \csname
  mn@eprint@\@tempb\endcsname \expandafter{\@tempc}}}

\bibitem[\protect\citeauthoryear{{Akaike}}{{Akaike}}{1974}]{Akaike1974ITAC...19..716A}
{Akaike} H.,  1974, IEEE Transactions on Automatic Control, \href
  {https://ui.adsabs.harvard.edu/abs/1974ITAC...19..716A} {19, 716}

\bibitem[\protect\citeauthoryear{{Ayres}, {Lyons}, {Ludwig}, {Caffau}  \&
  {Wedemeyer-B{\"o}hm}}{{Ayres} et~al.}{2013}]{Ayres2013ApJ...765...46A}
{Ayres} T.~R.,  {Lyons} J.~R.,  {Ludwig} H.~G.,  {Caffau} E.,
  {Wedemeyer-B{\"o}hm} S.,  2013, \mn@doi [\apj] {10.1088/0004-637X/765/1/46},
  \href {https://ui.adsabs.harvard.edu/abs/2013ApJ...765...46A} {765, 46}

\bibitem[\protect\citeauthoryear{{Bainbridge} \& {Webb}}{{Bainbridge} \&
  {Webb}}{2017a}]{Bainbridge_2017Univ....3...34B}
{Bainbridge} M.~B.,  {Webb} J.~K.,  2017a, \mn@doi [Universe]
  {10.3390/universe3020034}, \href
  {https://ui.adsabs.harvard.edu/abs/2017Univ....3...34B} {3, 34}

\bibitem[\protect\citeauthoryear{{Bainbridge} \& {Webb}}{{Bainbridge} \&
  {Webb}}{2017b}]{Bainbridge_2017MNRAS.468.1639B}
{Bainbridge} M.~B.,  {Webb} J.~K.,  2017b, \mn@doi [\mnras]
  {10.1093/mnras/stx179}, \href
  {https://ui.adsabs.harvard.edu/abs/2017MNRAS.468.1639B} {468, 1639}

\bibitem[\protect\citeauthoryear{{Baldwin}, {Burbidge}, {Hazard}, {Murdoch},
  {Robinson}  \& {Wampler}}{{Baldwin}
  et~al.}{1973}]{Baldwin1973ApJ...185..739B}
{Baldwin} J.~A.,  {Burbidge} E.~M.,  {Hazard} C.,  {Murdoch} H.~S.,  {Robinson}
  L.~B.,   {Wampler} E.~J.,  1973, \mn@doi [\apj] {10.1086/152451}, \href
  {https://ui.adsabs.harvard.edu/abs/1973ApJ...185..739B} {185, 739}

\bibitem[\protect\citeauthoryear{{Berg}, {Ellison}, {Prochaska}, {Venn}  \&
  {Dessauges-Zavadsky}}{{Berg} et~al.}{2015}]{Berg2015MNRAS.452.4326B}
{Berg} T. A.~M.,  {Ellison} S.~L.,  {Prochaska} J.~X.,  {Venn} K.~A.,
  {Dessauges-Zavadsky} M.,  2015, \mn@doi [\mnras] {10.1093/mnras/stv1577},
  \href {https://ui.adsabs.harvard.edu/abs/2015MNRAS.452.4326B} {452, 4326}

\bibitem[\protect\citeauthoryear{Blondel, Berthet, Cuturi, Frostig, Hoyer,
  Llinares-L{\'o}pez, Pedregosa  \& Vert}{Blondel
  et~al.}{2021}]{jaxopt_implicit_diff}
Blondel M.,  Berthet Q.,  Cuturi M.,  Frostig R.,  Hoyer S.,
  Llinares-L{\'o}pez F.,  Pedregosa F.,   Vert J.-P.,  2021, arXiv preprint
  arXiv:2105.15183

\bibitem[\protect\citeauthoryear{{Botelho}, {Milone}, {Mel{\'e}ndez},
  {Alves-Brito}, {Spina}  \& {Bean}}{{Botelho}
  et~al.}{2020}]{botelho2020MNRAS.499.2196B}
{Botelho} R.~B.,  {Milone} A. d.~C.,  {Mel{\'e}ndez} J.,  {Alves-Brito} A.,
  {Spina} L.,   {Bean} J.~L.,  2020, \mn@doi [\mnras] {10.1093/mnras/staa2917},
  \href {https://ui.adsabs.harvard.edu/abs/2020MNRAS.499.2196B} {499, 2196}

\bibitem[\protect\citeauthoryear{{Bozdogan}}{{Bozdogan}}{1987}]{Bozdogan1987}
{Bozdogan} H.,  1987, \mn@doi [Psychometrika] {10.1007/BF02294361}, 52, 345

\bibitem[\protect\citeauthoryear{Bradbury et~al.,}{Bradbury
  et~al.}{2018}]{jax2018github}
Bradbury J.,  et~al., 2018, {JAX}: composable transformations of
  {P}ython+{N}um{P}y programs, \url {http://github.com/google/jax}

\bibitem[\protect\citeauthoryear{Burnham \& Anderson}{Burnham \&
  Anderson}{2002}]{Burnham2002}
Burnham K.,  Anderson D.,  2002, Model selection and multimodel inference: a
  practical information-theoretic approach.
Springer Verlag, New York

\bibitem[\protect\citeauthoryear{{Carswell}}{{Carswell}}{2023}]{web:VPFIT}
{Carswell} R.~F.,  2023, \texttt{VPFIT} homepage:
  \url{https://people.ast.cam.ac.uk/~rfc/}, \url
  {https://people.ast.cam.ac.uk/~rfc/}

\bibitem[\protect\citeauthoryear{{Carswell} \& {Webb}}{{Carswell} \&
  {Webb}}{2014}]{Carswell_2014ascl.soft08015C}
{Carswell} R.~F.,  {Webb} J.~K.,  2014, {VPFIT: Voigt profile fitting program},
  {Astrophysics Source Code Library} (\mn@eprint {ascl} {1408.015})

\bibitem[\protect\citeauthoryear{{Carswell}, {Hilliard}, {Strittmatter},
  {Taylor}  \& {Weymann}}{{Carswell}
  et~al.}{1975}]{Carswell1975ApJ...196..351C}
{Carswell} R.~F.,  {Hilliard} R.~L.,  {Strittmatter} P.~A.,  {Taylor} D.~J.,
  {Weymann} R.~J.,  1975, \mn@doi [\apj] {10.1086/153418}, \href
  {https://ui.adsabs.harvard.edu/abs/1975ApJ...196..351C} {196, 351}

\bibitem[\protect\citeauthoryear{{Carswell}, {Jorgenson}, {Wolfe}  \&
  {Murphy}}{{Carswell} et~al.}{2011}]{Carswell2011}
{Carswell} R.~F.,  {Jorgenson} R.~A.,  {Wolfe} A.~M.,   {Murphy} M.~T.,  2011,
  \mn@doi [\mnras] {10.1111/j.1365-2966.2010.17854.x}, \href
  {https://ui.adsabs.harvard.edu/abs/2011MNRAS.411.2319C} {411, 2319}

\bibitem[\protect\citeauthoryear{{Caughlan}}{{Caughlan}}{1965}]{Caughlan1965}
{Caughlan} G.~R.,  1965, \mn@doi [\apj] {10.1086/148155}, \href
  {https://ui.adsabs.harvard.edu/abs/1965ApJ...141..688C} {141, 688}

\bibitem[\protect\citeauthoryear{{Chiappini}, {Ekstr{\"o}m}, {Meynet},
  {Hirschi}, {Maeder}  \& {Charbonnel}}{{Chiappini}
  et~al.}{2008}]{chiappini2008A&A...479L...9C}
{Chiappini} C.,  {Ekstr{\"o}m} S.,  {Meynet} G.,  {Hirschi} R.,  {Maeder} A.,
  {Charbonnel} C.,  2008, \mn@doi [\aap] {10.1051/0004-6361:20078698}, \href
  {https://ui.adsabs.harvard.edu/abs/2008A&A...479L...9C} {479, L9}

\bibitem[\protect\citeauthoryear{{Crossfield} et~al.,}{{Crossfield}
  et~al.}{2019}]{Crossfield2019ApJ...871L...3C}
{Crossfield} I.~J.~M.,  et~al., 2019, \mn@doi [\apjl]
  {10.3847/2041-8213/aaf9b6}, \href
  {https://ui.adsabs.harvard.edu/abs/2019ApJ...871L...3C} {871, L3}

\bibitem[\protect\citeauthoryear{{Cui}, {Bechtold}, {Ge}  \& {Meyer}}{{Cui}
  et~al.}{2005}]{Cui2005}
{Cui} J.,  {Bechtold} J.,  {Ge} J.,   {Meyer} D.~M.,  2005, \mn@doi [\apj]
  {10.1086/444368}, \href
  {https://ui.adsabs.harvard.edu/abs/2005ApJ...633..649C} {633, 649}

\bibitem[\protect\citeauthoryear{{Dekker}, {D'Odorico}, {Kaufer}, {Delabre}  \&
  {Kotzlowski}}{{Dekker} et~al.}{2000}]{Dekker2000SPIE.4008..534D}
{Dekker} H.,  {D'Odorico} S.,  {Kaufer} A.,  {Delabre} B.,   {Kotzlowski} H.,
  2000, in {Iye} M.,  {Moorwood} A.~F.,  eds,  Society of Photo-Optical
  Instrumentation Engineers (SPIE) Conference Series Vol. 4008, Optical and IR
  Telescope Instrumentation and Detectors. pp 534--545,
  \mn@doi{10.1117/12.395512}

\bibitem[\protect\citeauthoryear{{Fenner}, {Murphy}  \& {Gibson}}{{Fenner}
  et~al.}{2005}]{fenner2005MNRAS.358..468F}
{Fenner} Y.,  {Murphy} M.~T.,   {Gibson} B.~K.,  2005, \mn@doi [\mnras]
  {10.1111/j.1365-2966.2005.08781.x}, \href
  {https://ui.adsabs.harvard.edu/abs/2005MNRAS.358..468F} {358, 468}

\bibitem[\protect\citeauthoryear{Geyer}{Geyer}{2011}]{Geyer_introMCMC}
Geyer C.,  2011, Introduction to Markov Chain Monte Carlo.
CRC Press, pp 3--48, \mn@doi{10.1201/b10905-2}

\bibitem[\protect\citeauthoryear{{Goldbach} \& {Nollez}}{{Goldbach} \&
  {Nollez}}{1987}]{Goldbach_1987A&A...181..203G}
{Goldbach} C.,  {Nollez} G.,  1987, \aap, \href
  {https://ui.adsabs.harvard.edu/abs/1987A&A...181..203G} {181, 203}

\bibitem[\protect\citeauthoryear{{Goto} et~al.,}{{Goto}
  et~al.}{2003}]{Goto2003ApJ...598.1038G}
{Goto} M.,  et~al., 2003, \mn@doi [\apj] {10.1086/378978}, \href
  {https://ui.adsabs.harvard.edu/abs/2003ApJ...598.1038G} {598, 1038}

\bibitem[\protect\citeauthoryear{{Halfen}, {Woolf}  \& {Ziurys}}{{Halfen}
  et~al.}{2017}]{Halfen2017ApJ...845..158H}
{Halfen} D.~T.,  {Woolf} N.~J.,   {Ziurys} L.~M.,  2017, \mn@doi [\apj]
  {10.3847/1538-4357/aa816b}, \href
  {https://ui.adsabs.harvard.edu/abs/2017ApJ...845..158H} {845, 158}

\bibitem[\protect\citeauthoryear{{Haridass} \& {Huber}}{{Haridass} \&
  {Huber}}{1994}]{Haridass_1994ApJ...420..433H}
{Haridass} C.,  {Huber} K.~P.,  1994, \mn@doi [\apj] {10.1086/173573}, \href
  {https://ui.adsabs.harvard.edu/abs/1994ApJ...420..433H} {420, 433}

\bibitem[\protect\citeauthoryear{Harris et~al.,}{Harris
  et~al.}{2020}]{harris2020array_numpy}
Harris C.~R.,  et~al., 2020, \mn@doi [Nature] {10.1038/s41586-020-2649-2}, 585,
  357

\bibitem[\protect\citeauthoryear{{Henkel}, {Downes}, {Wei{\ss}}, {Riechers}  \&
  {Walter}}{{Henkel} et~al.}{2010}]{Henkel2010A&A...516A.111H}
{Henkel} C.,  {Downes} D.,  {Wei{\ss}} A.,  {Riechers} D.,   {Walter} F.,
  2010, \mn@doi [\aap] {10.1051/0004-6361/200912889}, \href
  {https://ui.adsabs.harvard.edu/abs/2010A&A...516A.111H} {516, A111}

\bibitem[\protect\citeauthoryear{{Henkel} et~al.,}{{Henkel}
  et~al.}{2014}]{Henkel2014A&A...565A...3H}
{Henkel} C.,  et~al., 2014, \mn@doi [\aap] {10.1051/0004-6361/201322962}, \href
  {https://ui.adsabs.harvard.edu/abs/2014A&A...565A...3H} {565, A3}

\bibitem[\protect\citeauthoryear{{Horne}}{{Horne}}{1986}]{Horne1986PASP...98..609H}
{Horne} K.,  1986, \mn@doi [\pasp] {10.1086/131801}, \href
  {https://ui.adsabs.harvard.edu/abs/1986PASP...98..609H} {98, 609}

\bibitem[\protect\citeauthoryear{Hunter}{Hunter}{2007}]{Hunter_2007Matplotlib}
Hunter J.~D.,  2007, \mn@doi [Computing in Science \& Engineering]
  {10.1109/MCSE.2007.55}, 9, 90

\bibitem[\protect\citeauthoryear{{Hurvich} \& {Tsai}}{{Hurvich} \&
  {Tsai}}{1989}]{Hurvich1989}
{Hurvich} C.~M.,  {Tsai} C.-L.,  1989, \mn@doi [Biometrika]
  {10.1093/biomet/76.2.297}, 76, 297

\bibitem[\protect\citeauthoryear{{Iben} \& {Renzini}}{{Iben} \&
  {Renzini}}{1984}]{iben1984PhR...105..329I}
{Iben} I.,  {Renzini} A.,  1984, \mn@doi [\physrep]
  {10.1016/0370-1573(84)90142-X}, \href
  {https://ui.adsabs.harvard.edu/abs/1984PhR...105..329I} {105, 329}

\bibitem[\protect\citeauthoryear{{Kimble} et~al.,}{{Kimble}
  et~al.}{1998}]{Kimble1998ApJ...492L..83K}
{Kimble} R.~A.,  et~al., 1998, \mn@doi [\apjl] {10.1086/311102}, \href
  {https://ui.adsabs.harvard.edu/abs/1998ApJ...492L..83K} {492, L83}

\bibitem[\protect\citeauthoryear{{Kobayashi}, {Karakas}  \&
  {Umeda}}{{Kobayashi} et~al.}{2011}]{kobayashi2011MNRAS.414.3231K}
{Kobayashi} C.,  {Karakas} A.~I.,   {Umeda} H.,  2011, \mn@doi [\mnras]
  {10.1111/j.1365-2966.2011.18621.x}, \href
  {https://ui.adsabs.harvard.edu/abs/2011MNRAS.414.3231K} {414, 3231}

\bibitem[\protect\citeauthoryear{{Kobayashi}, {Karakas}  \&
  {Lugaro}}{{Kobayashi} et~al.}{2020}]{Kobayashi2020ApJ...900..179K}
{Kobayashi} C.,  {Karakas} A.~I.,   {Lugaro} M.,  2020, \mn@doi [\apj]
  {10.3847/1538-4357/abae65}, \href
  {https://ui.adsabs.harvard.edu/abs/2020ApJ...900..179K} {900, 179}

\bibitem[\protect\citeauthoryear{{Lai}, {Ubachs}, {De Oliveira}  \&
  {Salumbides}}{{Lai} et~al.}{2020}]{Lai_2020Atoms...8...62L}
{Lai} K.-F.,  {Ubachs} W.,  {De Oliveira} N.,   {Salumbides} E.~J.,  2020,
  \mn@doi [Atoms] {10.3390/atoms8030062}, \href
  {https://ui.adsabs.harvard.edu/abs/2020Atoms...8...62L} {8, 62}

\bibitem[\protect\citeauthoryear{{Lambert} \& {Sneden}}{{Lambert} \&
  {Sneden}}{1977}]{lambert1977ApJ...215..597L}
{Lambert} D.~L.,  {Sneden} C.,  1977, \mn@doi [\apj] {10.1086/155393}, \href
  {https://ui.adsabs.harvard.edu/abs/1977ApJ...215..597L} {215, 597}

\bibitem[\protect\citeauthoryear{{Lee}, {Webb}, {Carswell}  \&
  {Milakovi{\'c}}}{{Lee} et~al.}{2021a}]{Lee2021aivpfit}
{Lee} C.-C.,  {Webb} J.~K.,  {Carswell} R.~F.,   {Milakovi{\'c}} D.,  2021a,
  \mn@doi [\mnras] {10.1093/mnras/stab977}, \href
  {https://ui.adsabs.harvard.edu/abs/2021MNRAS.504.1787L} {504, 1787}

\bibitem[\protect\citeauthoryear{{Lee}, {Webb}, {Milakovi{\'c}}  \&
  {Carswell}}{{Lee} et~al.}{2021b}]{Lee2021nonunique}
{Lee} C.-C.,  {Webb} J.~K.,  {Milakovi{\'c}} D.,   {Carswell} R.~F.,  2021b,
  \mn@doi [\mnras] {10.1093/mnras/stab2005}, \href
  {https://ui.adsabs.harvard.edu/abs/2021MNRAS.507...27L} {507, 27}

\bibitem[\protect\citeauthoryear{{Lee}, {Webb}, {Carswell}, {Dzuba}, {Flambaum}
   \& {Milakovi{\'c}}}{{Lee} et~al.}{2023}]{Lee2023}
{Lee} C.-C.,  {Webb} J.~K.,  {Carswell} R.~F.,  {Dzuba} V.~A.,  {Flambaum}
  V.~V.,   {Milakovi{\'c}} D.,  2023, \mn@doi [\mnras] {10.1093/mnras/stad600},
  \href {https://ui.adsabs.harvard.edu/abs/2023MNRAS.521..850L} {521, 850}

\bibitem[\protect\citeauthoryear{{Levshakov}, {Centuri{\'o}n}, {Molaro}  \&
  {Kostina}}{{Levshakov} et~al.}{2006}]{Levshakov2006A&A...447L..21L}
{Levshakov} S.~A.,  {Centuri{\'o}n} M.,  {Molaro} P.,   {Kostina} M.~V.,  2006,
  \mn@doi [\aap] {10.1051/0004-6361:200600001}, \href
  {https://ui.adsabs.harvard.edu/abs/2006A&A...447L..21L} {447, L21}

\bibitem[\protect\citeauthoryear{{Li}, {Lundberg}, {Berzinsh}, {Johansson}  \&
  {Svanberg}}{{Li} et~al.}{2000}]{Li_2000JPhB...33.5593L}
{Li} Z.~S.,  {Lundberg} H.,  {Berzinsh} U.,  {Johansson} S.,   {Svanberg} S.,
  2000, \mn@doi [Journal of Physics B Atomic Molecular Physics]
  {10.1088/0953-4075/33/24/311}, \href
  {https://ui.adsabs.harvard.edu/abs/2000JPhB...33.5593L} {33, 5593}

\bibitem[\protect\citeauthoryear{{Limongi} \& {Chieffi}}{{Limongi} \&
  {Chieffi}}{2018}]{Limongi2018}
{Limongi} M.,  {Chieffi} A.,  2018, \mn@doi [\apjs] {10.3847/1538-4365/aacb24},
  \href {https://ui.adsabs.harvard.edu/abs/2018ApJS..237...13L} {237, 13}

\bibitem[\protect\citeauthoryear{{Marconi} et~al.,}{{Marconi}
  et~al.}{2022}]{Marconi_2022SPIE12184E..24M}
{Marconi} A.,  et~al., 2022, in {Evans} C.~J.,  {Bryant} J.~J.,   {Motohara}
  K.,  eds,  Society of Photo-Optical Instrumentation Engineers (SPIE)
  Conference Series Vol. 12184, Ground-based and Airborne Instrumentation for
  Astronomy IX. p. 1218424, \mn@doi{10.1117/12.2628689}

\bibitem[\protect\citeauthoryear{{Marconi} et~al.,}{{Marconi}
  et~al.}{2024}]{Marconi_2024arXiv240714601M}
{Marconi} A.,  et~al., 2024, \mn@doi [arXiv e-prints]
  {10.48550/arXiv.2407.14601}, \href
  {https://ui.adsabs.harvard.edu/abs/2024arXiv240714601M} {p. arXiv:2407.14601}

\bibitem[\protect\citeauthoryear{{Meyer}, {York}, {Black}, {Chaffee}  \&
  {Foltz}}{{Meyer} et~al.}{1986}]{Meyer1986}
{Meyer} D.~M.,  {York} D.~G.,  {Black} J.~H.,  {Chaffee} F.~H. J.,   {Foltz}
  C.~B.,  1986, \mn@doi [\apjl] {10.1086/184739}, \href
  {https://ui.adsabs.harvard.edu/abs/1986ApJ...308L..37M} {308, L37}

\bibitem[\protect\citeauthoryear{{Meynet}, {Ekstr{\"o}m}  \& {Maeder}}{{Meynet}
  et~al.}{2006}]{Meynet2006A}
{Meynet} G.,  {Ekstr{\"o}m} S.,   {Maeder} A.,  2006, \mn@doi [\aap]
  {10.1051/0004-6361:20053070}, \href
  {https://ui.adsabs.harvard.edu/abs/2006A&A...447..623M} {447, 623}

\bibitem[\protect\citeauthoryear{Modigliani, Sosnowska  \& Lovis}{Modigliani
  et~al.}{2023}]{ESPRESSO_DRS3.0.0}
Modigliani A.,  Sosnowska D.,   Lovis C.,  2023, ESPRESSO Pipeline User Manual
  version 3.0.0.
\url
  {https://ftp.eso.org/pub/dfs/pipelines/instruments/espresso/espdr-pipeline-manual-3.0.0.pdf}

\bibitem[\protect\citeauthoryear{{Molaro}}{{Molaro}}{2003}]{molaro2003ASPC..304..221M}
{Molaro} P.,  2003, in {Charbonnel} C.,  {Schaerer} D.,   {Meynet} G.,  eds,
  Astronomical Society of the Pacific Conference Series Vol. 304, CNO in the
  Universe. p.~221

\bibitem[\protect\citeauthoryear{{Molaro} et~al.,}{{Molaro}
  et~al.}{2023}]{Molaro2023A&A...679A..72M}
{Molaro} P.,  et~al., 2023, \mn@doi [\aap] {10.1051/0004-6361/202347676}, \href
  {https://ui.adsabs.harvard.edu/abs/2023A&A...679A..72M} {679, A72}

\bibitem[\protect\citeauthoryear{{Muller}, {Gu{\'e}lin}, {Dumke}, {Lucas}  \&
  {Combes}}{{Muller} et~al.}{2006}]{Muller2006}
{Muller} S.,  {Gu{\'e}lin} M.,  {Dumke} M.,  {Lucas} R.,   {Combes} F.,  2006,
  \mn@doi [\aap] {10.1051/0004-6361:20065187}, \href
  {https://ui.adsabs.harvard.edu/abs/2006A&A...458..417M} {458, 417}

\bibitem[\protect\citeauthoryear{{Murphy}}{{Murphy}}{2018}]{Murphy2018_uves_popler_zndo...1297190M}
{Murphy} M.,  2018, {MTMurphy77/UVES\_popler: UVES\_popler: POst-PipeLine
  Echelle Reduction software}, \mn@doi{10.5281/zenodo.1297190}

\bibitem[\protect\citeauthoryear{{Murphy}, {Kacprzak}, {Savorgnan}  \&
  {Carswell}}{{Murphy} et~al.}{2019}]{Murphy_2019MNRAS.482.3458M}
{Murphy} M.~T.,  {Kacprzak} G.~G.,  {Savorgnan} G. A.~D.,   {Carswell} R.~F.,
  2019, \mn@doi [\mnras] {10.1093/mnras/sty2834}, \href
  {https://ui.adsabs.harvard.edu/abs/2019MNRAS.482.3458M} {482, 3458}

\bibitem[\protect\citeauthoryear{{Murphy} et~al.,}{{Murphy}
  et~al.}{2022}]{Murphy2022A&A...658A.123M}
{Murphy} M.~T.,  et~al., 2022, \mn@doi [\aap] {10.1051/0004-6361/202142257},
  \href {https://ui.adsabs.harvard.edu/abs/2022A&A...658A.123M} {658, A123}

\bibitem[\protect\citeauthoryear{{Noterdaeme} et~al.,}{{Noterdaeme}
  et~al.}{2017}]{Noterdaeme2017A&A...597A..82N}
{Noterdaeme} P.,  et~al., 2017, \mn@doi [\aap] {10.1051/0004-6361/201629173},
  \href {https://ui.adsabs.harvard.edu/abs/2017A&A...597A..82N} {597, A82}

\bibitem[\protect\citeauthoryear{{Pasquini} \& {Milakovi{\'c}}}{{Pasquini} \&
  {Milakovi{\'c}}}{2024}]{Pasquini_2024arXiv240514955P}
{Pasquini} L.,  {Milakovi{\'c}} D.,  2024, \mn@doi [arXiv e-prints]
  {10.48550/arXiv.2405.14955}, \href
  {https://ui.adsabs.harvard.edu/abs/2024arXiv240514955P} {p. arXiv:2405.14955}

\bibitem[\protect\citeauthoryear{{Pepe} et~al.,}{{Pepe}
  et~al.}{2021}]{Pepe2021A&A...645A..96P}
{Pepe} F.,  et~al., 2021, \mn@doi [\aap] {10.1051/0004-6361/202038306}, \href
  {https://ui.adsabs.harvard.edu/abs/2021A&A...645A..96P} {645, A96}

\bibitem[\protect\citeauthoryear{Phan, Pradhan  \& Jankowiak}{Phan
  et~al.}{2019}]{Phan2019_numpyro}
Phan D.,  Pradhan N.,   Jankowiak M.,  2019, arXiv preprint arXiv:1912.11554

\bibitem[\protect\citeauthoryear{{Prantzos}, {Aubert}  \& {Audouze}}{{Prantzos}
  et~al.}{1996}]{Prantzos1996A&A...309..760P}
{Prantzos} N.,  {Aubert} O.,   {Audouze} J.,  1996, \aap, \href
  {https://ui.adsabs.harvard.edu/abs/1996A&A...309..760P} {309, 760}

\bibitem[\protect\citeauthoryear{{Robertson}}{{Robertson}}{1986}]{Robertson_1986PASP...98.1220R}
{Robertson} J.~G.,  1986, \mn@doi [\pasp] {10.1086/131925}, \href
  {https://ui.adsabs.harvard.edu/abs/1986PASP...98.1220R} {98, 1220}

\bibitem[\protect\citeauthoryear{{Romano}}{{Romano}}{2022}]{Romano2022A&ARv..30....7R}
{Romano} D.,  2022, \mn@doi [\aapr] {10.1007/s00159-022-00144-z}, \href
  {https://ui.adsabs.harvard.edu/abs/2022A&ARv..30....7R} {30, 7}

\bibitem[\protect\citeauthoryear{{Romano} \& {Matteucci}}{{Romano} \&
  {Matteucci}}{2003}]{Romano2003MNRAS.342..185R}
{Romano} D.,  {Matteucci} F.,  2003, \mn@doi [\mnras]
  {10.1046/j.1365-8711.2003.06526.x}, \href
  {https://ui.adsabs.harvard.edu/abs/2003MNRAS.342..185R} {342, 185}

\bibitem[\protect\citeauthoryear{{Schmidt} et~al.,}{{Schmidt}
  et~al.}{2021}]{Schmidt2021A&A...646A.144S}
{Schmidt} T.~M.,  et~al., 2021, \mn@doi [\aap] {10.1051/0004-6361/202039345},
  \href {https://ui.adsabs.harvard.edu/abs/2021A&A...646A.144S} {646, A144}

\bibitem[\protect\citeauthoryear{{Songaila} et~al.,}{{Songaila}
  et~al.}{1994}]{Songaila1994}
{Songaila} A.,  et~al., 1994, \mn@doi [\nat] {10.1038/371043a0}, \href
  {https://ui.adsabs.harvard.edu/abs/1994Natur.371...43S} {371, 43}

\bibitem[\protect\citeauthoryear{{Spite} et~al.,}{{Spite}
  et~al.}{2005}]{spite05}
{Spite} M.,  et~al., 2005, \mn@doi [\aap] {10.1051/0004-6361:20041274}, \href
  {https://ui.adsabs.harvard.edu/abs/2005A&A...430..655S} {430, 655}

\bibitem[\protect\citeauthoryear{{Spite}, {Spite}  \& {Barbuy}}{{Spite}
  et~al.}{2021}]{spite2021A&A...652A..97S}
{Spite} M.,  {Spite} F.,   {Barbuy} B.,  2021, \mn@doi [\aap]
  {10.1051/0004-6361/202141741}, \href
  {https://ui.adsabs.harvard.edu/abs/2021A&A...652A..97S} {652, A97}

\bibitem[\protect\citeauthoryear{{Strittmatter}, {Carswell}, {Burbidge},
  {Hazard}, {Baldwin}, {Robinson}  \& {Wampler}}{{Strittmatter}
  et~al.}{1973}]{Strittmatter1973ApJ...183..767S}
{Strittmatter} P.~A.,  {Carswell} R.~F.,  {Burbidge} E.~M.,  {Hazard} C.,
  {Baldwin} J.~A.,  {Robinson} L.,   {Wampler} E.~J.,  1973, \mn@doi [\apj]
  {10.1086/152265}, \href
  {https://ui.adsabs.harvard.edu/abs/1973ApJ...183..767S} {183, 767}

\bibitem[\protect\citeauthoryear{{Szentgyorgyi} et~al.,}{{Szentgyorgyi}
  et~al.}{2018}]{Szentgyorgyi_2018SPIE10702E..1RS}
{Szentgyorgyi} A.,  et~al., 2018, in {Evans} C.~J.,  {Simard} L.,   {Takami}
  H.,  eds,  Society of Photo-Optical Instrumentation Engineers (SPIE)
  Conference Series Vol. 10702, Ground-based and Airborne Instrumentation for
  Astronomy VII. p. 107021R, \mn@doi{10.1117/12.2313539}

\bibitem[\protect\citeauthoryear{Vehtari, Gelman, Simpson, Carpenter  \&
  B{\"u}rkner}{Vehtari et~al.}{2021}]{Rhat}
Vehtari A.,  Gelman A.,  Simpson D.,  Carpenter B.,   B{\"u}rkner P.-C.,  2021,
  \mn@doi [Bayesian Analysis] {10.1214/20-BA1221}, 16, 667

\bibitem[\protect\citeauthoryear{Virtanen et~al.,}{Virtanen
  et~al.}{2020}]{2020SciPy-NMeth}
Virtanen P.,  et~al., 2020, \mn@doi [Nature Methods]
  {10.1038/s41592-019-0686-2}, \href {https://rdcu.be/b08Wh} {17, 261}

\bibitem[\protect\citeauthoryear{{Vogt} et~al.,}{{Vogt}
  et~al.}{1994}]{Vogt1994SPIE.2198..362V}
{Vogt} S.~S.,  et~al., 1994, in {Crawford} D.~L.,  {Craine} E.~R.,  eds,
  Society of Photo-Optical Instrumentation Engineers (SPIE) Conference Series
  Vol. 2198, Instrumentation in Astronomy VIII. p.~362,
  \mn@doi{10.1117/12.176725}

\bibitem[\protect\citeauthoryear{{Wallstr{\"o}m}, {Muller}  \&
  {Gu{\'e}lin}}{{Wallstr{\"o}m} et~al.}{2016}]{Wallstrom2016A&A...595A..96W}
{Wallstr{\"o}m} S.~H.~J.,  {Muller} S.,   {Gu{\'e}lin} M.,  2016, \mn@doi
  [\aap] {10.1051/0004-6361/201628615}, \href
  {https://ui.adsabs.harvard.edu/abs/2016A&A...595A..96W} {595, A96}

\bibitem[\protect\citeauthoryear{{Webb}, {Lee}, {Carswell}  \&
  {Milakovi{\'c}}}{{Webb} et~al.}{2021}]{Webb2021}
{Webb} J.~K.,  {Lee} C.-C.,  {Carswell} R.~F.,   {Milakovi{\'c}} D.,  2021,
  \mn@doi [\mnras] {10.1093/mnras/staa3551}, \href
  {https://ui.adsabs.harvard.edu/abs/2021MNRAS.501.2268W} {501, 2268}

\bibitem[\protect\citeauthoryear{{Webb}, {Lee}  \& {Milakovi{\'c}}}{{Webb}
  et~al.}{2022}]{Webb2022}
{Webb} J.~K.,  {Lee} C.-C.,   {Milakovi{\'c}} D.,  2022, \mn@doi [Universe]
  {10.3390/universe8050266}, \href
  {https://ui.adsabs.harvard.edu/abs/2022Univ....8..266W} {8, 266}

\bibitem[\protect\citeauthoryear{{Welsh}, {Cooke}, {Fumagalli}  \&
  {Pettini}}{{Welsh} et~al.}{2020}]{Welsh_2020MNRAS.494.1411W}
{Welsh} L.,  {Cooke} R.,  {Fumagalli} M.,   {Pettini} M.,  2020, \mn@doi
  [\mnras] {10.1093/mnras/staa807}, \href
  {https://ui.adsabs.harvard.edu/abs/2020MNRAS.494.1411W} {494, 1411}

\bibitem[\protect\citeauthoryear{{Wiescher}, {G{\"o}rres}, {Uberseder},
  {Imbriani}  \& {Pignatari}}{{Wiescher}
  et~al.}{2010}]{Weischer2010ARNPS..60..381W}
{Wiescher} M.,  {G{\"o}rres} J.,  {Uberseder} E.,  {Imbriani} G.,   {Pignatari}
  M.,  2010, \mn@doi [Annual Review of Nuclear and Particle Science]
  {10.1146/annurev.nucl.012809.104505}, \href
  {https://ui.adsabs.harvard.edu/abs/2010ARNPS..60..381W} {60, 381}

\bibitem[\protect\citeauthoryear{{Yan} et~al.,}{{Yan}
  et~al.}{2023}]{yan2023A&A...670A..98Y}
{Yan} Y.~T.,  et~al., 2023, \mn@doi [\aap] {10.1051/0004-6361/202244584}, \href
  {https://ui.adsabs.harvard.edu/abs/2023A&A...670A..98Y} {670, A98}

\bibitem[\protect\citeauthoryear{{Zafar}, {Centuri{\'o}n}, {P{\'e}roux},
  {Molaro}, {D'Odorico}, {Vladilo}  \& {Popping}}{{Zafar}
  et~al.}{2014}]{zafar2014MNRAS.444..744Z}
{Zafar} T.,  {Centuri{\'o}n} M.,  {P{\'e}roux} C.,  {Molaro} P.,  {D'Odorico}
  V.,  {Vladilo} G.,   {Popping} A.,  2014, \mn@doi [\mnras]
  {10.1093/mnras/stu1473}, \href
  {https://ui.adsabs.harvard.edu/abs/2014MNRAS.444..744Z} {444, 744}

\bibitem[\protect\citeauthoryear{{Zechmeister}, {Anglada-Escud{\'e}}  \&
  {Reiners}}{{Zechmeister} et~al.}{2014}]{Zechmeister_2014A&A...561A..59Z}
{Zechmeister} M.,  {Anglada-Escud{\'e}} G.,   {Reiners} A.,  2014, \mn@doi
  [\aap] {10.1051/0004-6361/201322746}, \href
  {https://ui.adsabs.harvard.edu/abs/2014A&A...561A..59Z} {561, A59}

\makeatother
\end{thebibliography}




\appendix

\section{Example models}\label{sec:example_models}
\begin{figure*}
    \centering
    \includegraphics[trim=0pt 1.2cm 0pt 0cm, clip, width=0.94\textwidth]{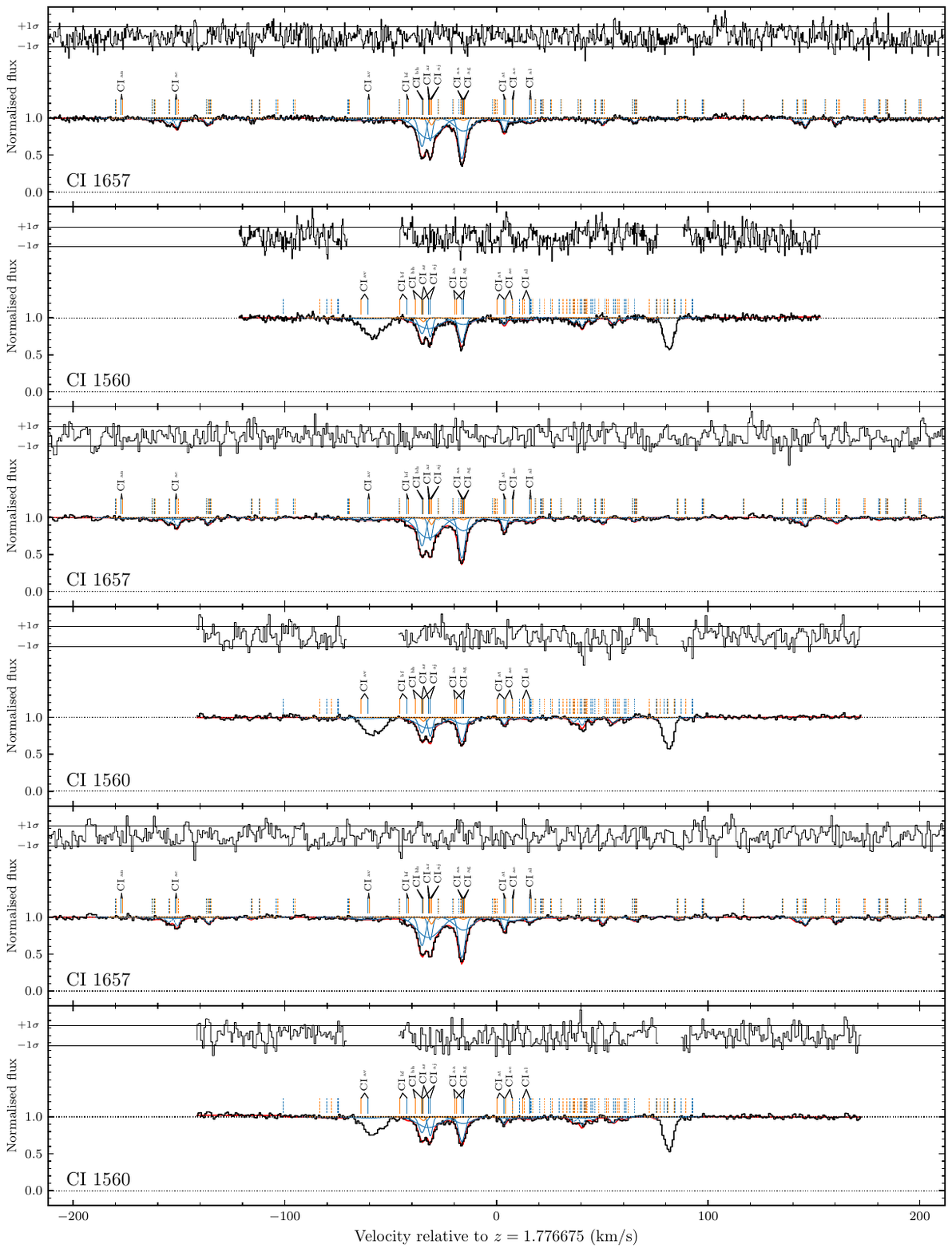}
    \caption{The same as \figref{fig:spectrum} but spectra of the three epochs are shown separately. First two rows are epoch I, second two rows are epoch II, and the final two rows are epoch III.  $\chi^2_\nu=0.6362$.}
    \label{fig:model_cratio28p14_separated}
\end{figure*}

\begin{figure*}
    \centering
    \includegraphics[trim=0pt 1.2cm 0pt 0cm, clip, width=0.94\textwidth]{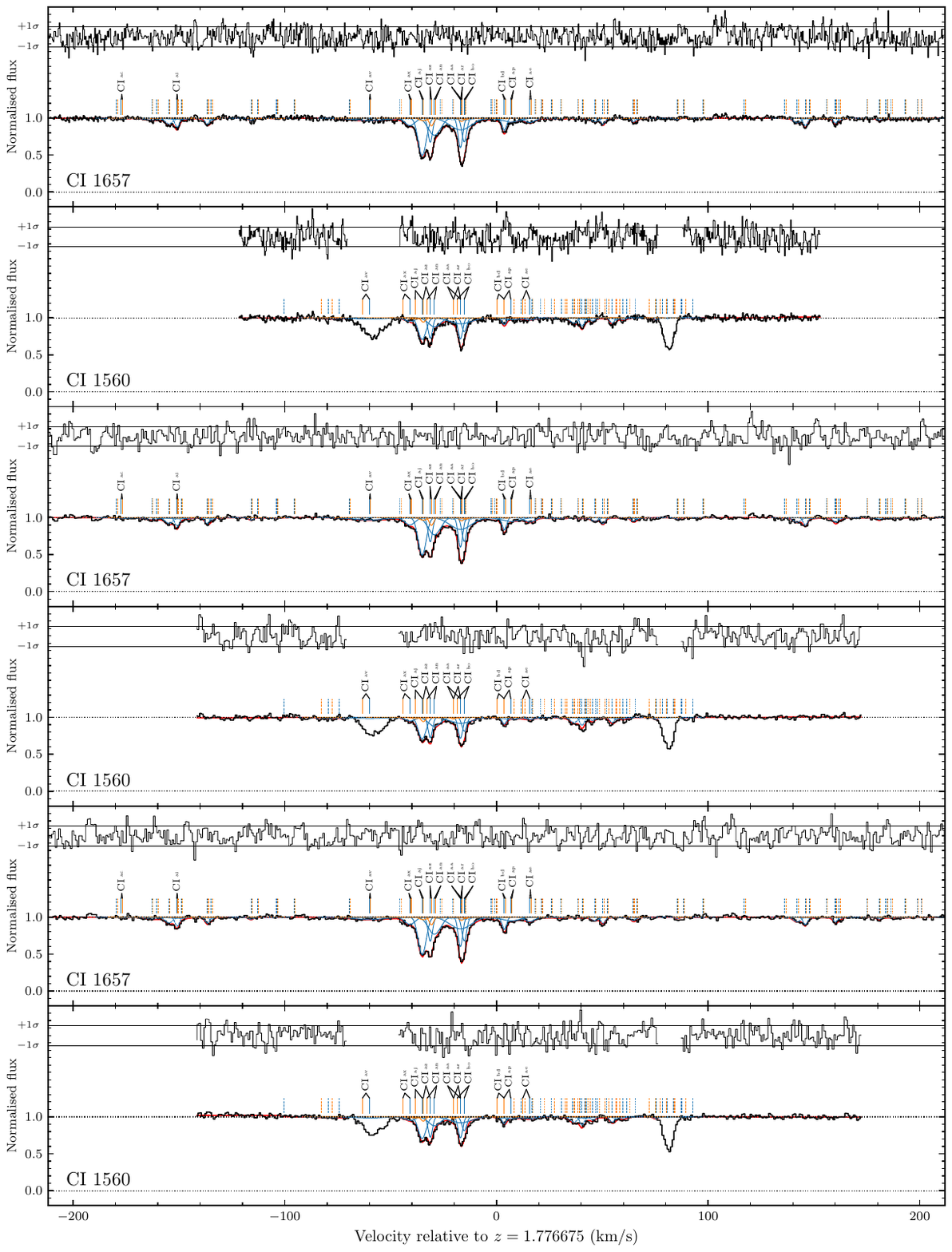}
    \caption{The same as \figref{fig:model_cratio28p14_separated} but for $\cratiomath=20$. $\chi^2_\nu = 0.6269$.}
    \label{fig:model_cratio20}
\end{figure*}

\begin{figure*}
    \centering
    \includegraphics[trim=0pt 1.2cm 0pt 0cm, clip, width=0.94\textwidth]{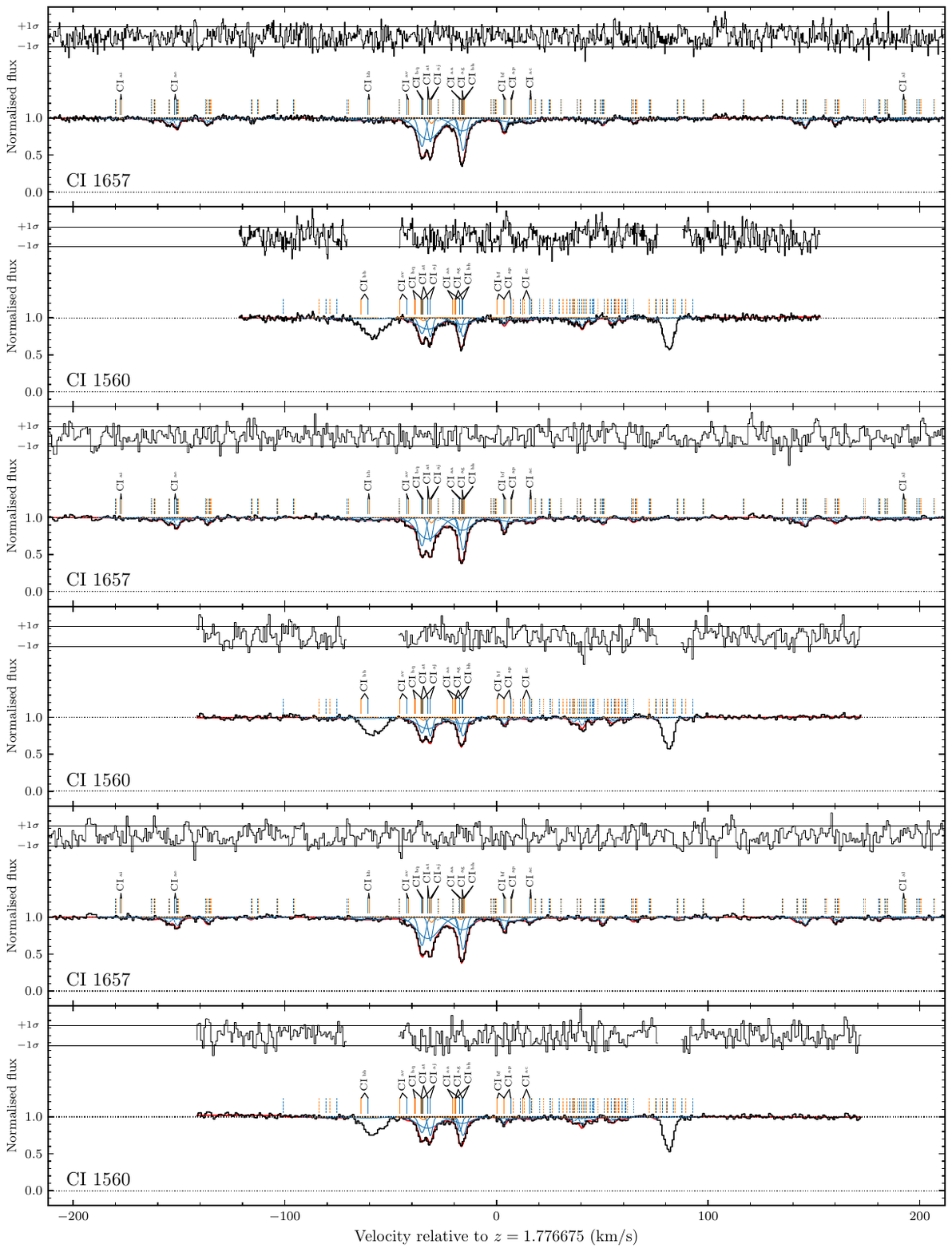}
    \caption{The same as \figref{fig:model_cratio28p14_separated} but for $\cratiomath=90.71$. $\chi^2_\nu = 0.6216$.}
    \label{fig:model_cratio91}
\end{figure*}

\begin{figure*}
    \centering
    \includegraphics[trim=0pt 1.2cm 0pt 0cm, clip, width=0.94\textwidth]{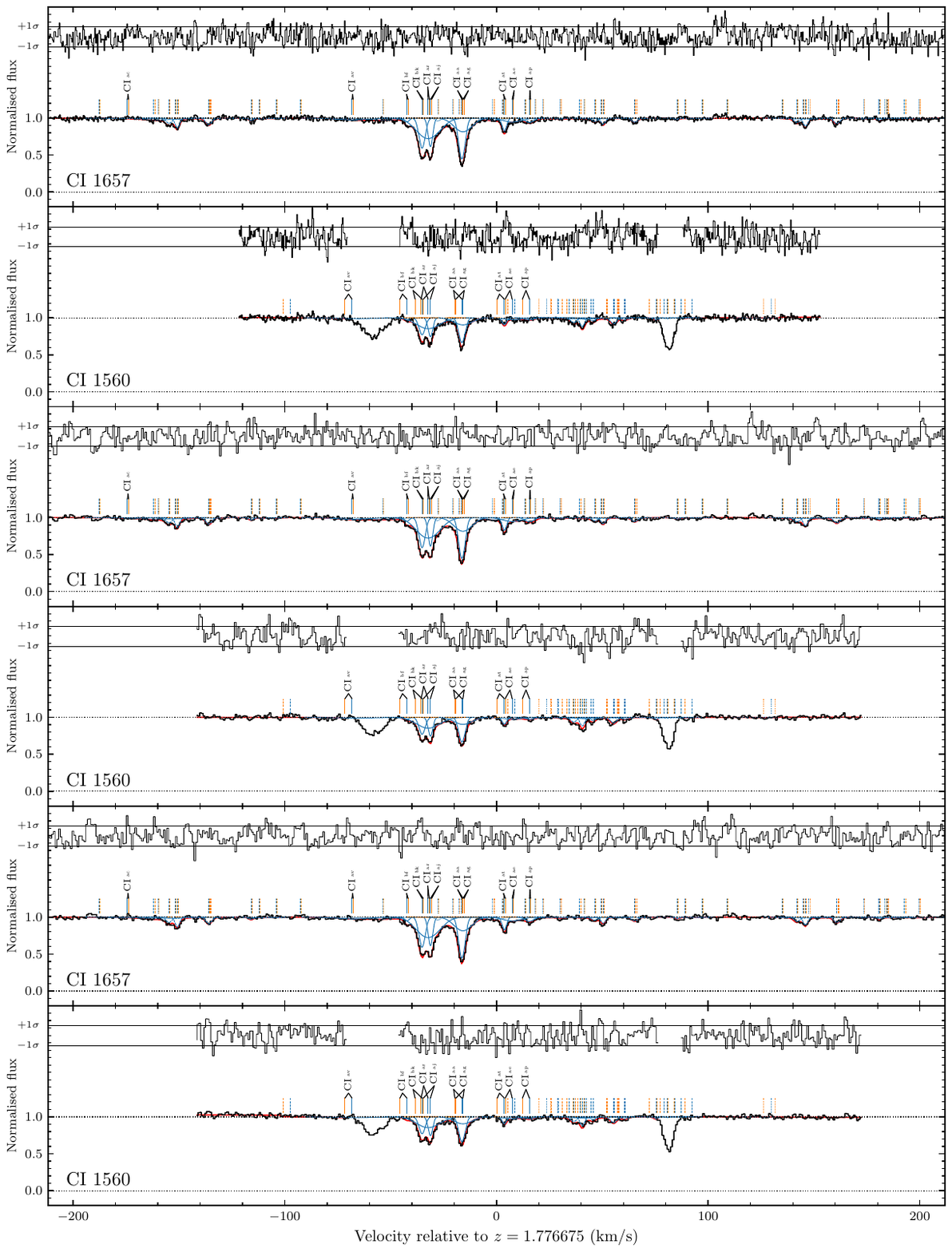}
    \caption{The same as \figref{fig:model_cratio28p14_separated} but for $\cratiomath=500$. $\chi^2_\nu = 0.6448$.}
    \label{fig:model_cratio500}
\end{figure*}

The figures in this Appendix show four example {\aivpfit} models, each with a different {\cratio}, derived as a part of the analysis presented in the main text. Figure \ref{fig:model_cratio28p14_separated} shows the same model as is shown in \figref{fig:spectrum} ($\cratiomath=28.14$), but with spectra separated by epoch. The remaining Figures show models with $\cratiomath=20$ (\figref{fig:model_cratio20}), solar {\cratio} (=91, \figref{fig:model_cratio91}), and $\cratiomath=500$ (\figref{fig:model_cratio500}).

\FloatBarrier

\section{Comparing ESPRESSO C{\sc i} velocity structure with independent models from H$_2$}\label{sec:theoretical_models}

\citet{Cui2005} analysed H$_2$ rotational levels seen in the \emph{Hubble Space Telescope}/STIS \citep{Kimble1998ApJ...492L..83K} data associated with the DLA studied here. Assuming a single velocity component for H$_2$, a population analysis of its rotational levels yielded a H$_2$ gas temperature of $T=\qty{152(10)}{\kelvin}$. Subsequently, \citet{Carswell2011} examined the physical properties of the gas in more detail by combining the same STIS observations of H$_2$ with VLT/UVES \citep{Dekker2000SPIE.4008..534D} and Keck/HIRES \citep{Vogt1994SPIE.2198..362V} observations covering the wavelength ranges containing heavy element absorption. In the \citet{Carswell2011} analysis, 52 H$_2$ rotational levels seen in the STIS data were modelled together with five \atom{C}{i} transitions from UVES and HIRES, assuming three velocity components for both H$_2$ and \atom{C}{i}. Only two components had reliably measured temperatures, $T({\rm H}_2) = 86^{+14}_{-10}\qty{}{\kelvin}$, and $177^{+30}_{-22}\qty{}{\kelvin}$ (at $\zabs=1.7763702$ and $1.7765246$, i.e. their components N1 and N2, respectively). A lower limit of $T({\rm H}_2) \gtrsim \qty{200}{\kelvin}$ was obtained for a component at $\zabs=1.7767176$ (their N3$^\prime$). \citet{Carswell2011} suggested N3$^\prime$ to be a blend of several more narrow features that are unresolved in their data. They also suggested that N2 is mostly thermally broadened whereas N1 and N3$^\prime$ are dominated by turbulent motions. The best estimate for the kinetic temperature of component N2 provided by \citet{Carswell2011} is $T=\qty{218}{\kelvin}$, from \atom{C}{i} line broadening. 

\begin{figure*}
    \includegraphics[width=\textwidth]{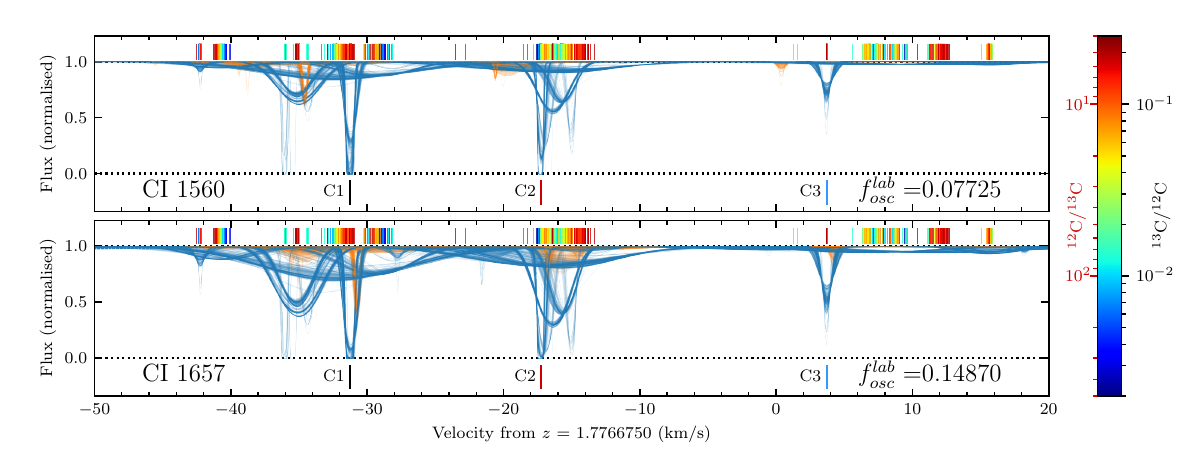}
    \caption{Theoretical absorption system models before convolution with the assumed IPs. The models and the colour coding are the same as in \figref{fig:nonuniqueness}. A single {\aivpfit} model may contain between zero and five narrow ($b<\kmps{0.55}$) velocity components, with most models containing two of them. The three most prominent narrow components are at \kmps{-31.2} (C1), \kmps{-17.2} (C2), and \kmps{+3.7} (C3). The redshifts and temperatures of C1 and C2 are in agreement with literature reported redshifts of H$_2$ absorption and their temperatures \citep{Carswell2011}, but {\aivpfit} derived them completely independently and without instructions. The presence of C3 was previously unknown. 
    }
    \label{fig:theoretical_models}
\end{figure*}

Out of the 285 surviving {\aivpfit} models produced in this work, 281 contain \atom{C}{i} components with $b$-parameters consistent with temperatures $T \leqslant \qty{218}{\kelvin}$. 26 models have one, 129 models have two, 112 models have three, 11 models have four, and three models have five such components. \figref{fig:theoretical_models} shows where those components tend to fall within the absorption complex. The three most prominent narrow components are found at $\zabs=1.7763856$ (C1, at \kmps{-31.2} in \figref{fig:theoretical_models}), $\zabs=1.7765156$ (C2, at \kmps{-17.2}), and $\zabs=1.7767093$ (C3, at \kmps{+3.7}). The redshifts of the C1 and C2 are \kmps{1.7} and \kmps{1.0} away from the redshifts of N1 and N2 reported by \citet{Carswell2011}. 
Such shifts are not surprising, considering that ESPRESSO data allowed for resolving more structure compared to previous UVES and HIRES observations ($R=42000$ and 48000, respectively, compared to ESPRESSO's $R=140000$). Component C3 was not known before.

We next derived the temperatures of C1, C2, and C3 assuming that their broadening is only due to thermal gas motions. In doing that, we considered components that are $\leq\pm\kmps{0.5}$ away from their redshifts (see above), and have $b$-parameters corresponding to a maximum {\ctwo} gas temperature of \qty{1000}{\kelvin} ($b=\kmps{1.18}$). Median temperatures obtained this way are $T({\rm C1})=51^{+124}_{-27}\qty{}{\kelvin}$, $T({\rm C2})=60^{+12}_{-48}\qty{}{\kelvin}$, and $T({\rm C3})=180^{+159}_{-50}\qty{}{\kelvin}$. The quoted errors correspond to the 16\textsuperscript{th} and the 84\textsuperscript{th} percentiles. 

Identifying our C1 with N1 from \citet{Carswell2011}, we found the temperature derived from the ESPRESSO \atom{C}{i} line widths ($T=51^{+124}_{-27}\qty{}{\kelvin}$) to be in agreement with the temperature derived from H$_2$ by \citet{Carswell2011} ($T=86^{+14}_{-10}\qty{}{\kelvin}$). While it may appear that temperature obtained by us for C2 ($T=60^{+12}_{-48}\qty{}{\kelvin}$) is in disagreement with the temperature \citet{Carswell2011} reported for N2 ($T=177^{+30}_{-22}\qty{}{\kelvin}$), this is not the case, as ESPRESSO data revealed the presence of several components not resolved in the lower resolution UVES and HIRES data used by \citet{Carswell2011}. Additional ESPRESSO observations may be used to confirm the presence of cold components at \kmps{-42}, \kmps{-36}, \kmps{-34}, and \kmps{-15} (with respect to $z=1.7766750$) contained in some {\aivpfit} models. Observations using even higher spectral resolution spectrographs \citep[such as G-CLEF, with a planned $R=\qty{300000}{}$,][]{Szentgyorgyi_2018SPIE10702E..1RS}, would also be useful. 

Parameter constraints in our calculations required $b<\kmps{10}$, an empirically-guided value. We note that some models reveal the presence of broad components i.e.\ with $b$-parameters near or at the upper limit. It is unavoidably the case that there may be clumps of blended lines that cannot be resolved (given the intrinsic line properties and spectral resolution). The number of free parameters in our models are decided on by an information criterion, giving the benefit of reproducibility. Given the constraints imposed, statistically the data do not justify additional parameters. Examining where components with $b\geq\kmps{8}$ fall within the absorption complex, we found them to be concentrated at the following locations: at \kmps{-175} (14\% of all broad components), \kmps{-87} (1\%), \kmps{-72} (2\%), \kmps{-64} (44\%), \kmps{-34} (1\%), \kmps{-16} ($\leq1\%$), \kmps{+9} (8\%), \kmps{+13} (15\%), and \kmps{+194} (15\%). Two of the strong \atom{C}{i} components from which {\cratio} constraint is derived are C1 and C2 (see \figref{fig:theoretical_models}), i.e.\ the overwhelming majority ($>98\%$) of the broad components are located far from regions providing constraints on {\cratio}. 

\FloatBarrier
\section{MCMC posteriors}\label{sec:mcmc}

\begin{figure*}
    \centering
    \includegraphics[scale=0.9]{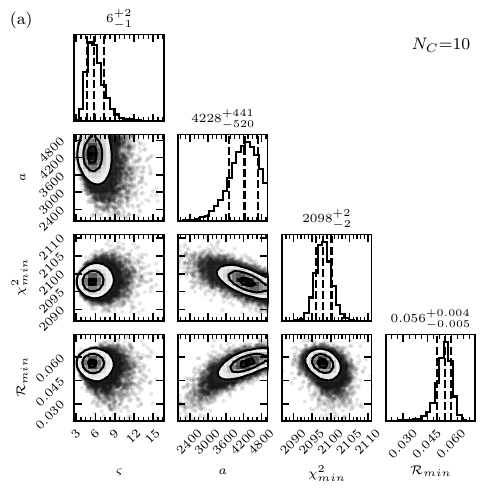}
    \includegraphics[scale=0.9]{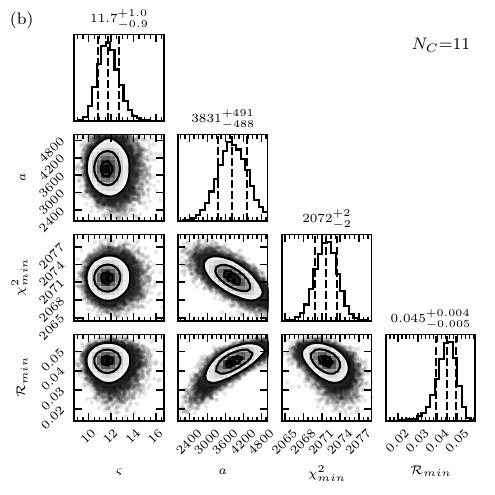}
    \includegraphics[scale=0.9]{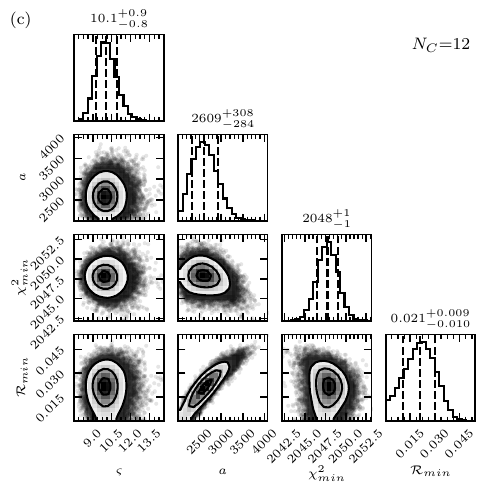}
    \includegraphics[scale=0.9]{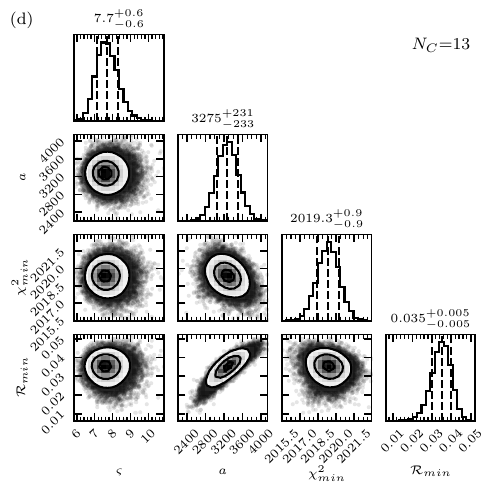}
    \includegraphics[scale=0.9]{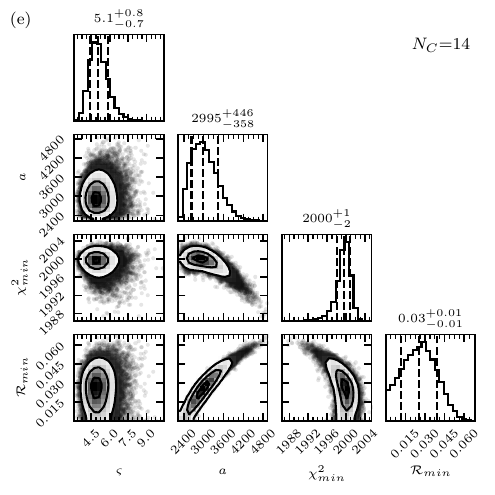}
    \caption{Posterior distributions and covariances of Equation \eqref{eq:parabola} parameters after MCMC optimisation for the five data subsets (panels a--e). $N_{C}$ for each subset is indicated in the top right corner of each panel. Numbers are the median and the central 68\% limits, indicated by the dashed lines. }
    \label{fig:corner_plots}
\end{figure*}

\figref{fig:corner_plots} shows the marginalised posterior distributions for parameters included in MCMC calculations presented in \secref{sec:analysis}, and their mutual covariances. 

\bsp	
\label{lastpage}
\end{document}